\documentclass[letterpaper,english,prx,twocolumn,amsmath,amssymb,superscriptaddress]{revtex4-2}
\usepackage[T1]{fontenc}
\usepackage{color}
\usepackage{babel}
\usepackage{prettyref}
\usepackage{dsfont}
\usepackage{amsmath}
\usepackage{amssymb}
\usepackage{graphicx}
\usepackage{wasysym}
\usepackage[unicode=true,
 bookmarks=true,bookmarksnumbered=false,bookmarksopen=false,
 breaklinks=false,pdfborder={0 0 1},backref=false,colorlinks=false]
 {hyperref}
\hypersetup{pdftitle={Furutsu-Novikov--like cross-corelation--response relations for systems driven by shot noise},
 pdfauthor={Jakob Stubenrauch, Benjamin Lindner},
 colorlinks=true,linkcolor=black,citecolor=black,urlcolor=black,filecolor=black}

\makeatletter

\usepackage{tikz}
\usetikzlibrary{calc}

\usepackage[latin1]{inputenc}

\newrefformat{eq}{\hyperref[#1]{Eq.~(\ref{#1})}}
\newrefformat{cap}{\hyperref[#1]{Fig.~\ref{#1}}}
\newrefformat{fig}{\hyperref[#1]{Fig.~\ref{#1}}}
\newrefformat{tab}{\hyperref[#1]{Table ~\ref{#1}}}
\newrefformat{sec}{\hyperref[#1]{Sec.~\ref{#1}}}
\newrefformat{subsec}{\hyperref[#1]{Sec.~\ref{#1}}}
\newrefformat{sub}{\hyperref[#1]{Sec.~\ref{#1}}}
\newrefformat{cha}{\hyperref[#1]{Chapter~\ref{#1}}}
\newrefformat{appsec}{\hyperref[#1]{App.~\ref{#1}}}

\usepackage{siunitx}
\usepackage{MnSymbol}
\DeclareMathAlphabet\mathbb{U}{msb}{m}{n}
\newcommand\hemdash{\hbox{---}\kern-.5em---}

\makeatother

\begin{document}
\global\long\def\T{\mathrm{T}}%
\global\long\def\cO{\mathcal{O}}%
\global\long\def\cF{\mathcal{F}}%
\global\long\def\cI{\mathcal{I}}%
\global\long\def\epsilon{\varepsilon}%
\global\long\def\cD{\mathcal{D}}%
\global\long\def\cZ{\mathcal{Z}}%
\global\long\def\bu{\boldsymbol{u}}%
\global\long\def\pprime{\prime\prime}%
\global\long\def\ppp{\prime\prime\prime}%
\global\long\def\Qty#1#2{\qty{#1}{#2}}%
\global\long\def\He{\text{He}}%

\title{Furutsu-Novikov--like cross-correlation--response relations \\
for systems driven by shot noise}
\author{Jakob Stubenrauch}
\email{Corresponding author. {jakob.stubenrauch@rwth-aachen.de}}

\affiliation{Bernstein Center for Computational Neuroscience Berlin, Philippstra{\ss}e
13, Haus 2, 10115 Berlin, Germany}
\affiliation{Physics Department of Humboldt University Berlin, Newtonstra{\ss}e
15, 12489 Berlin, Germany}
\author{Benjamin Lindner}
\affiliation{Bernstein Center for Computational Neuroscience Berlin, Philippstra{\ss}e
13, Haus 2, 10115 Berlin, Germany}
\affiliation{Physics Department of Humboldt University Berlin, Newtonstra{\ss}e
15, 12489 Berlin, Germany}
\date{\today}
\begin{abstract}
We consider a dynamic system that is driven by an intensity-modulated
Poisson process with intensity $\Lambda(t)=\lambda(t)+\varepsilon\nu(t)$.
We derive an exact relation between the input-output cross-correlation
in the spontaneous state ($\varepsilon=0$) and the linear response
to the modulation ($\varepsilon>0$). If $\epsilon$ is sufficiently
small, linear response theory captures the full response. The relation
can be regarded as a variant of the Furutsu-Novikov theorem for the
case of shot noise. As we show, the relation is still valid in the
presence of additional independent noise. Furthermore, we derive an
extension to Cox-process input, which provides an instance of colored
shot noise. We discuss applications to particle detection and to neuroscience.
Using the new relation, we obtain a fluctuation-response-relation
for a leaky integrate-and-fire neuron. We also show how the new relation
can be used in a remote control problem in a recurrent neural network.
The relations are numerically tested for both stationary and non-stationary
dynamics. Lastly, extensions to marked Poisson processes and to higher-order
statistics are presented.

\end{abstract}
\maketitle

\section{Introduction}

Numerous systems in nature generate random sequences of events, which
in turn drive other systems. For instance, the photocurrent in a
detector \citep{gardiner1985handbook}, the spontaneous vacuum current
in an electrode \citep{Schottky1918_567}, neural firing driven by
other neurons' spikes \citep{gerstner2014neuronal}, and an economy
subject to the effect of seismic events \citep{Tesfamariam_Goda_Handbook_of_seismic_risk}
are all examples for systems driven by random point processes. 

A first approach to quantify the relation of a driving process and
an observable of the system is to compute the cross-correlation function.
Cross-correlations quantify the similarity of fluctuations in the
drive and the observable, yet they cannot be straightforwardly used
to predict the response to systematic perturbations of the drive.
The latter can be quantified by linear-response functions. 

If the driving process is colored Gaussian noise, the two important
statistics, cross-correlation functions and linear-response functions,
are linearly related by the Furutsu-Novikov theorem (FNT) \citep{Furutsu63_303,Novikov65_1290},
also known as Gaussian integration by parts \citep{Frisch1995turbulence}.
The relation has been frequently applied to study wave propagation
in random media \citep{Tarasov2023_214202}, turbulent flow \citep{Frisch1995turbulence,Krommes2002_352,Bentkamp2022_2088},
neural systems \citep{Lindner22_198101,Clark_23_118401,Puttkammer2024_0770},
and general stochastic processes \citep{Fox1986_476,hanggi1994colored,MarPug08,Athanassoulis2019_115217}.
In the neural context, the FNT is particularly interesting: In the
absence of an external stimulus, neurons in the cortex are not silent
but spike spontaneously \citep{MarBre02,DooDor16}; when a weak stimulus
is applied, the neurons respond linearly \citep{TchMal11,DooDor16,KoeGie08,BiaRie91,Kni72}.
As one of us showed \citep{Lindner22_198101} using the FNT, spontaneous
fluctuations and linear response can be connected in a fluctuation-response
relation (FRR). Yet, since the FNT relies on Gaussian noise, shot-noise--driven
systems, such as the systems mentioned above are not captured by this
approach.
\begin{figure}
\includegraphics{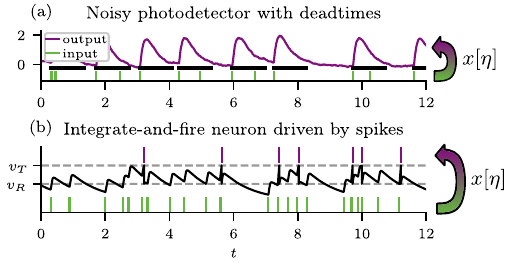}

\caption{\label{fig:models}Examples of trajectories of shot-noise--driven
systems. (a) Output (purple line) of a nonparalyzable photodetector
with deadtimes (black bars) driven by shot noise (green lines). (b)
Voltage trace (black), output spikes (purple), and input spikes (green)
for a leaky integrate-and-fire neuron \prettyref{eq:model}.}
\end{figure}

Modeling shot noise as Gaussian noise, known as the diffusion approximation
\citep{gerstner2014neuronal}, is only justified due to the central
limit theorem if the intensity of events is high and their amplitude
is low. However, for pyramidal neurons, as few as two input spikes
can be sufficient to trigger an output spike \citep{Markram1997_440,Badel2008_666},
which renders results based on the diffusion approximation inaccurate
\citep{Richardson10_178102}. Also for the broad problem of particle
detection, the shot-noise character of the input cannot be neglected
in many cases of interest. To extend the theory of stochastic processes
from a Gaussian description to a true shot-noise description, an important
stepping stone is to find an analogue of the FNT for shot-noise--driven
systems. Such an analogue must attribute input-output cross-correlations
of a shot-noise--driven black box to the response functions of the
black box.

In this paper we first consider an arbitrary system driven by an
inhomogeneous Poisson process, and derive a relation between its spontaneous
input-output cross-correlation and its response to a time-dependent
modulation of the input intensity. This cross-correlation--response
relation (CRR) can be regarded as a shot-noise analogue of the original
FNT. Although all necessary tools to do so are known \citep{last_penrose_2017},
such a relation has to our knowledge not been reported yet. We then
discuss the impact of additional noise and derive an extension to
the case where the input is correlated shot noise. Next, we test the
CRR for a minimal model of a particle detector and for a leaky integrate-and-fire
neuron. For the leaky integrate-and-fire neuron, we leverage the CRR
and the method of \citep{Lindner22_198101} to derive an FRR between
the spontaneous \emph{output} fluctuations of a shot-noise--driven
neuron and its systematic response to a time-dependent intensity modulation.
We numerically test the CRR and the FRR in the commonly studied stationary
case but demonstrate that the validity of the CRR extends to a non-stationary
scenario. Furthermore, we show how the CRR can be applied to the problem
of remote control in a recurrent neural network. Lastly, we present
extensions to Poissonian input with random amplitudes and to nonlinear
response functions.

\section{Cross-correlation--response relation}

We denote a system's observable $x[\eta(\circ);t]$ by a functional
of the entire input process $\eta(\circ)$. Both, the intrinsic time
argument $\circ$ and the functional dependence on $\eta$, will be
hidden, whenever not explicitly required, to ease the notation. The
additional scalar dependence on the observation time $t$ must reflect
causality. A simple example is linearly filtered shot noise, $x_{\kappa}(t)=\int^{t}dt^{\prime}\,\kappa(t-t^{\prime})\eta(t^{\prime})$;
however, the following applies to general functionals.

We consider inhomogeneous Poissonian input 
\begin{equation}
\eta(t)=\sum_{i}\delta(t-t_{i}),
\end{equation}
where $t_{i}$ are Poisson events with intensity $\Lambda(t)=\lambda(t)+\epsilon\nu(t)$,
$\lambda(t)$ is the baseline, and $\epsilon\nu(t)$ is a modulation.
We denote the observable's average over spontaneous ($\epsilon=0$)
realizations of the input by $\left\langle x(t)\right\rangle _{0}$,
thus the cross-correlation function between the input events and the
output without modulation is
\begin{equation}
C_{x\eta}(t,t^{\prime})=\left\langle x(t)\eta(t^{\prime})\right\rangle _{0}-\left\langle x(t)\right\rangle _{0}\left\langle \eta(t^{\prime})\right\rangle _{0}.\label{eq:cross_correlations}
\end{equation}
If we switch on the modulation, the average effect on the observable
can be represented perturbatively by a linear-response function $K$,
\begin{equation}
\left\langle x(t)\right\rangle _{\epsilon}=\left\langle x(t)\right\rangle _{0}+\epsilon\int dt^{\prime}\,K(t,t^{\prime})\nu(t^{\prime})+\cO(\epsilon^{2}).\label{eq:lin_resp_def}
\end{equation}
The $\cO(\epsilon^{2})$ corrections can be omitted for a sufficiently
weak modulation, and the linear-response function is given by a
functional derivative \citep{volterra1959theory}
\begin{equation}
K(t,t^{\prime})=\left.\frac{\delta}{\delta\Lambda(t^{\prime})}\left\langle x(t)\right\rangle _{0}\right|_{\Lambda=\lambda},\label{eq:lin_resp_constr}
\end{equation}
where
\begin{equation}
\frac{\delta}{\delta f(t)}g[f]\equiv\left.\frac{d}{dh}g[f(\circ)+h\delta(t-\circ)]\right|_{h=0}
\end{equation}
captures the infinitesimal change of a functional due to an infinitesimal
perturbation at time $t$.

In the following, we express both the response- and the cross-correlation
functions by the characteristic functional of the input process which
leads us to the CRR. Specifically, we represent $x[\eta;t]$ by its
functional Fourier transform with respect to $\eta$, 
\begin{equation}
x[\eta;t]=\int\cD u\,y[u;t]e^{iu^{\T}\eta},\label{eq:funct_ft}
\end{equation}
where $u^{\T}\eta\equiv\int dt\,u(t)\eta(t)$ and $\int\cD u\equiv\lim_{M\rightarrow\infty}\int_{-\infty}^{\infty}du_{1}...\int_{-\infty}^{\infty}du_{M}$
with $u_{i}=u(t_{i})$, and $\{t_{1},...,t_{M}\}$ is an equidistant
discretization of the interval $[0,T]$. For instance, an exponentiated
linear filter $x_{\beta}[\eta;t]=\exp\left(\int^{t}dt^{\prime}\beta(t-t^{\prime})\eta(t^{\prime})\right)$
has the functional Fourier transform $y[u;t]=\delta\left[iu(\circ)-\beta(t-\circ)\Theta(t-\circ)\right]$
with the functional Dirac delta $\delta[f(\circ)]\equiv\lim_{M\rightarrow\infty}\prod_{i=1}^{M}\delta(f(t_{i}))$
and the Heaviside function $\Theta(t)$. Thus, in general, the functional
Fourier transform dissects functional maps $\eta\mapsto x[\eta;t]$
into a linear combination of exponentiated linear filters. This is
useful as it establishes a link to characteristic functionals, as
used below. The formal limit $M\rightarrow\infty$ assumes convergence
of $x$ upon decreasing the time step $T/M$, which is also a requirement
for Euler integration. For finite $M$, \prettyref{eq:funct_ft} is
the usual $M$-dimensional Fourier transform. For a formal classification
of the existence of the Fourier transform of functionals of Poisson
processes, see \citep[Lemma 18.4]{last_penrose_2017}.

Plugging \prettyref{eq:funct_ft} into \prettyref{eq:lin_resp_constr}
yields
\begin{equation}
\begin{aligned}K(t,t^{\prime}) & =\int\cD u\,y[u;t]\,\left.\frac{\delta}{\delta\Lambda(t^{\prime})}\left\langle e^{iu^{\T}\eta}\right\rangle _{0}\right|_{\Lambda=\lambda}\\
 & =\int\cD u\,y[u;t]e^{iu(t^{\prime})}Z[u]-\left\langle x(t)\right\rangle _{0},
\end{aligned}
\label{eq:lin_resp_by_char}
\end{equation}
where the characteristic functional of a Poisson process of intensity
$\lambda(t)$ is  \citep{Stratonovich67} 
\begin{equation}
Z[u]\equiv\left\langle e^{iu^{\T}\eta}\right\rangle _{0}=e^{\int\lambda(t)\left[e^{iu(t)}-1\right]dt}.
\end{equation}

The cross-correlation \prettyref{eq:cross_correlations} can be expressed
by a functional derivative of $Z[u]$ as well
\begin{equation}
\begin{aligned}C_{x\eta}(t,t^{\prime}) & =\int\cD u\,y[u;t]\frac{\delta}{\delta iu(t^{\prime})}Z[u]-\lambda(t^{\prime})\left\langle x(t)\right\rangle _{0}\\
=\lambda(t^{\prime}) & \left[\int\cD u\,y[u;t]Z[u]e^{iu(t^{\prime})}-\left\langle x(t)\right\rangle _{0}\right].
\end{aligned}
\label{eq:cross_corr_from_char_func}
\end{equation}
Comparing this result with \prettyref{eq:lin_resp_by_char}, we infer
for arbitrary systems driven by Poisson noise a CRR
\begin{equation}
C_{x\eta}(t,t^{\prime})=\lambda(t^{\prime})K(t,t^{\prime}).\label{eq:crr}
\end{equation}
Thus, irrespective of the details of the system, the input-output
cross-correlation, a measure of the spontaneous fluctuations, can
be fully attributed to the linear-response function, which determines
the leading order deviations \emph{away from} the spontaneous mean.
In particular, the cross-correlation is unaffected by nonlinear response
functions. \prettyref{eq:crr} should be put in context with the original
FNT for Gaussian noise $\xi$ \citep{Furutsu63_303,Novikov65_1290}
\begin{equation}
C_{x\xi}(t,t^{\prime\prime})=\int dt^{\prime}C_{\xi}(t^{\prime\prime},t^{\prime})K_{x\xi}(t,t^{\prime}),\label{eq:fnt_orig}
\end{equation}
where $C_{\xi}$ is the noise autocorrelation function and $K_{x\xi}$
is the response of $\left\langle x\right\rangle $ to a modulation
of the input mean $m(t)=\left\langle \xi(t)\right\rangle $. Remarkably,
for white Gaussian noise $C_{\xi}(t,t^{\prime})=\lambda(t^{\prime})\delta(t-t^{\prime})$
{[}recall that Poissonian noise is white, $C_{\eta}(t,t^{\prime})=\lambda(t^{\prime})\delta(t-t^{\prime})${]},
\prettyref{eq:fnt_orig} looks like \prettyref{eq:crr}, if we naively
identify $K$ with $K_{x\xi}$. However, there are important differences
between the two response functions. $K_{x\xi}$ is the response function
of the time-dependent mean output, $\left\langle x(t)\right\rangle $,
with respect to a modulation of the mean input in the presence of
Gaussian noise. The response function $K$ considers the same output
statistics $\left\langle x(t)\right\rangle $ but with respect to
a modulation of the intensity (the rate) of the driving Poisson process,
a modulation that affects not only the mean value but all higher cumulants
of the driving noise and in particular the noise intensity. An additional
difference is that $K$ is also shaped by the type of background noise,
which is here a Poisson process and not Gaussian noise. Hence, there
are two differences and in general \prettyref{eq:crr} and \prettyref{eq:fnt_orig}
constitute relations between input-output cross-correlations and response
functions for two distinct settings (different background noise) and
two different types of response functions. Nevertheless, as we show
in \prettyref{appsec:CRR-implies-white-noise}, the two relations
become equivalent in a diffusion limit.

Returning to the CRR \prettyref{eq:crr}, for a constant baseline
intensity $\lambda(t)\equiv\lambda_{0}$ and stationary dynamics of
the driven system, we may define $C_{x\eta}(\tau)\equiv C_{x\eta}(t+\tau,t)$
and $K(\tau)\equiv K(t+\tau,t)$ such that in frequency space  we
have
\begin{equation}
S_{x\eta}(\omega)=\lambda_{0}\chi(\omega),\label{eq:crr_stationary}
\end{equation}
where the cross-spectrum $S_{x\eta}(\omega)=\cF[C_{x\eta}](\omega)$
and susceptibility $\chi(\omega)=\cF[K](\omega)$ are the Fourier
transforms $\cF[f](\omega)=\int dt\,e^{i\omega t}f(t)$ of $C_{x\eta}$
and $K$ respectively.

For the linearly filtered shot noise $x_{\kappa}$ introduced above,
\prettyref{eq:crr} can be checked directly: Using the statistics
of Poisson processes \citep{Stratonovich67} one gets $C_{x_{\kappa}\eta}(t,t^{\prime})=\lambda(t^{\prime})\kappa(t-t^{\prime})\Theta(t-t^{\prime})$
and $K(t,t^{\prime})=\kappa(t-t^{\prime})\Theta(t-t^{\prime})$, which
confirms the CRR. In \prettyref{sec:Applications}, we demonstrate
the validity of \prettyref{eq:crr} for systems for which the explicit
computation of the statistics of interest is not feasible. Next,
we present two useful extensions of the CRR.

\subsection{Extension: additional noise\label{subsec:Additional-noise}}

Here we consider the prevalent situation in which the shot-noise--driven
system receives additional noise, for instance thermal noise or the
input of other random forces. The system's output at observation time
$t$ may be an arbitrary functional $\hat{x}[\eta,\xi;t]$ of both,
the Poissonian input $\eta$ and the additional random force $\xi$.
Assuming statistical independence of $\eta$ and $\xi$, the $\xi$-averaged
cross-correlation between the Poissonian drive $\eta$ and the output
$\hat{x}$ is
\begin{align}
\left\langle C_{\hat{x}\eta}(t,t^{\prime})\right\rangle _{\xi} & \equiv\left\langle \left\langle \hat{x}[\eta,\xi;t]\eta(t^{\prime})\right\rangle _{\eta}-\left\langle \hat{x}[\eta,\xi;t]\right\rangle _{\eta}\left\langle \eta(t^{\prime})\right\rangle _{\eta}\right\rangle _{\xi}\nonumber \\
 & =\left\langle \left\langle \hat{x}[\eta,\xi;t]\right\rangle _{\xi}\eta(t^{\prime})\right\rangle _{\eta}-\left\langle \left\langle \hat{x}[\eta,\xi;t]\right\rangle _{\xi}\right\rangle _{\eta}\left\langle \eta(t^{\prime})\right\rangle _{\eta}\nonumber \\
 & \equiv C_{\left\langle \hat{x}\right\rangle _{\xi}\eta}(t,t^{\prime}).
\end{align}
This is equal to the cross-correlation between the Poissonian drive
$\eta$ and the $\xi$-averaged output $\left\langle \hat{x}\right\rangle _{\xi}$.
Furthermore, we may interchange the order of $\xi$-averaging and
applying a linear differential operator, thus the $\xi$-averaged
response of $\hat{x}$ to modulations of the intensity of $\eta$
is
\begin{equation}
\left\langle \left.\frac{\delta}{\delta\Lambda(t^{\prime})}\left\langle \hat{x}(t)\right\rangle _{\eta}\right|_{\Lambda=\lambda}\right\rangle _{\xi}=\left.\frac{\delta}{\delta\Lambda(t^{\prime})}\left\langle \left\langle \hat{x}(t)\right\rangle _{\xi}\right\rangle _{\eta}\right|_{\Lambda=\lambda}.
\end{equation}
When defining $x[\eta;t]\equiv\left\langle \hat{x}[\eta,\xi;t]\right\rangle _{\xi}$,
we are back at the case without additional noise discussed above.
Thus, for additional noise sources that are independent of the drive
$\eta$, the CRR \prettyref{eq:crr} is still valid. For an experimenter
this means that, for instance, uncontrolled thermal fluctuations are
not detrimental for the CRR and will simply average out. This fact
is exploited in \prettyref{subsec:detector}.

\subsection{Extension: shot noise with temporal correlations\label{subsec:cox}}

The Poisson process considered so far is white noise $C_{\eta}(t,t^{\prime})\propto\delta(t-t^{\prime})$,
and we assumed knowledge of $\lambda(t)$. Both assumptions can for
instance be problematic when studying neural networks. A useful noise
model solving both problems is the Cox process, or doubly stochastic
Poisson process, where events are conditionally Poissonian with intensity
$\lambda(t)$, but $\lambda(t)$ is itself a random process. Here,
we choose the intensity $\lambda(t)=\Theta[\phi(t)]\phi(t)$ where
$\phi(t)$ is a Gaussian process and we assume that  the mean $m(t)$
and autocorrelation $C_{\phi}(t,t^{\prime})$ are chosen such that
$\phi(t)<0$ only rarely and we may set $\lambda(t)=\phi(t)$. The
limits of validity of this choice are discussed in \prettyref{appsec:Non-negative-cox}.
The cumulants of the Cox process are 
\begin{equation}
\left\langle \eta(t)\right\rangle =m(t),\,\,\,\,C_{\eta}(t,t^{\prime})=C_{\phi}(t,t^{\prime})+m(t)\delta(t-t^{\prime}),\label{eq:cox_cumulants}
\end{equation}
the latter reflecting the law of total variance, thus the process
is indeed temporally correlated (colored) noise. Next, we derive a
somewhat involved extension of the CRR for this case. However, as
we show afterwards, the involved expression can be simplified under
reasonable assumptions.

\subsubsection{Exact CRR for systems driven by a Gaussian Cox process}

The characteristic functional of a Cox process with a Gaussian-process
intensity is given by $Z_{m,C_{\phi}}[u]=\left\langle Z_{\phi}[u]\right\rangle _{\phi\sim\mathcal{N}(m,C_{\phi})}$
\citep{Bartlett1963_296} where $Z_{\phi}[u]=\exp\left(\int\phi(t)\left[e^{iu(t)}-1\right]dt\right)$
is the characteristic functional of the Poisson process and the average
is taken over the ensemble of Gaussian processes with mean $m$ and
autocorrelation $C_{\phi}$, thus
\begin{equation}
Z_{m,C_{\phi}}[u]=e^{m^{\T}\left(e^{iu}-1\right)+\frac{1}{2}\left(e^{iu}-1\right)^{\T}C_{\phi}\left(e^{iu}-1\right)},\label{eq:char_func_cox}
\end{equation}
where 
\begin{equation}
\begin{aligned}f^{\T}g & =\int dt\,f(t)g(t),\\
\frac{1}{2}f^{\T}Cg & =\int dt_{1}dt_{2}\frac{1}{2}f(t_{1})C(t_{1},t_{2})g(t_{2}).
\end{aligned}
\end{equation}
To decompose the cross-correlation into linear-response functions,
we need to consider the linear response 
\begin{equation}
K_{xm}(t,t^{\prime})=\left.\frac{\delta}{\delta M(t^{\prime})}\left\langle x(t)\right\rangle _{M,C_{\phi}}\right|_{M=m}
\end{equation}
to mean modulations where the subscripts of the expectation value
denote the noise parameters. Additionally, we need to take into account
the linear response $K_{xC_{\phi}}(t,t^{\prime},t^{\pprime})$ to
autocorrelation modulations 
\begin{equation}
C_{\phi}(t^{\prime},t^{\pprime})\rightarrow C_{\phi}(t^{\prime},t^{\pprime})+\epsilon D_{\phi}(t^{\prime},t^{\pprime}).
\end{equation}
This response function can be defined by
\begin{equation}
\begin{aligned}\left\langle x(t)\right\rangle _{m,C_{\phi}+\epsilon D} & =\left\langle x(t)\right\rangle _{m,C_{\phi}}\\
+\epsilon\int dt^{\prime}dt^{\pprime} & D_{\phi}(t^{\prime},t^{\pprime})K_{xC_{\phi}}(t,t^{\prime},t^{\pprime})+\cO(\epsilon^{2})
\end{aligned}
\end{equation}
and $K_{xC_{\phi}}(t,t^{\prime},t^{\pprime})=\left.\frac{\delta}{\delta C(t^{\prime},t^{\pprime})}\left\langle x(t)\right\rangle _{m,C}\right|_{C=C_{\phi}}$
can be computed using the two-point functional derivative 
\begin{equation}
\frac{\delta}{\delta g(t,t^{\prime})}f[g]\equiv\left.\frac{\partial}{\partial h}f[g+h\delta(\circ_{1}-t)\delta(\circ_{2}-t^{\prime})]\right|_{h=0}
\end{equation}
where $\circ_{1/2}$ denote the two intrinsic time arguments of $g$.
Proceeding analogously to the case of Poissonian input leads to a
CRR for Cox-process--driven systems
\begin{equation}
\begin{aligned}C_{x\eta} & (t,t^{\prime})=m(t^{\prime})K_{xm}(t,t^{\prime})\\
+ & \int dt^{\pprime}C_{\phi}(t^{\prime},t^{\pprime})\left[2K_{xC_{\phi}}(t,t^{\prime},t^{\pprime})+K_{xm}(t,t^{\pprime})\right]
\end{aligned}
\label{eq:cox_crr-3}
\end{equation}
which may alternatively be written as {[}see \prettyref{eq:cox_cumulants}{]}
\begin{equation}
\begin{aligned}C_{x\eta}(t,t^{\prime})= & \int dt^{\pprime}[C_{\phi}(t^{\prime},t^{\pprime})2K_{xC_{\phi}}(t,t^{\prime},t^{\pprime})\\
 & +C_{\eta}(t^{\prime},t^{\pprime})K_{xm}(t,t^{\pprime})].
\end{aligned}
\end{equation}
This formulation reveals a more complicated relation between the
input-output cross-correlation and the response statistics than in
the simpler Poissonian and Gaussian cases, Eqs. \eqref{eq:crr} and
\eqref{eq:fnt_orig}, respectively, because here the right-hand
side involves different statistics of the input noise as well as response
functions to modulations of different parameters. Note that in the
limit $C_{\phi}(t,t^{\prime})\rightarrow0$, in which the intensity
becomes deterministic $\left\langle \left[\lambda(t)-m(t)\right]^{2}\right\rangle \rightarrow0$,
\prettyref{eq:crr} is recovered with $m(t)\hat{=}\lambda(t)$ and
$K_{xm}\hat{=}K$. For the linear filter, \prettyref{eq:cox_crr-3}
can be checked directly, here $K_{xC_{\phi}}\equiv0$ (see \prettyref{appsec:Non-negative-cox}).

To compute the correlation response $K_{xC_{\phi}}$ numerically,
one must modulate the autocorrelation function of a Gaussian process.
To illustrate a way to do this, consider the Langevin equation
\begin{equation}
\tau_{\phi}\dot{\phi}=-\phi+\sqrt{2\sigma^{2}\tau_{\phi}[1+\epsilon s(t)]}\xi(t),
\end{equation}
where $\xi$ is centered Gaussian white noise. The autocorrelation
of $\phi$ is then for $\tau>0$
\begin{equation}
\begin{aligned}C_{\phi}(t+\tau,t) & =\sigma^{2}e^{-\tau/\tau_{\phi}}\\
+\epsilon\frac{2\sigma^{2}}{\tau_{\phi}} & e^{-\tau/\tau_{\phi}}\int_{0}^{\infty}d\Delta\,e^{-2\Delta/\tau_{\phi}}s(t-\Delta).
\end{aligned}
\label{eq:generate_acf_modulation}
\end{equation}
If the spontaneous autocorrelation takes the form $\sigma^{2}e^{-\tau/\tau_{\phi}}$,
one can thus generate the modulated process by tuning $s$ such that
the second line in \prettyref{eq:generate_acf_modulation} is the
desired $\epsilon D(t+\tau,t)$, the response to which defines the
function $K_{xC_{\phi}}$. More general spontaneous situations can
be achieved analogously by higher dimensional Markovian embedding.

\subsubsection{Weakly correlation-responsive regime for systems driven by a Gaussian
Cox process}

For systems that respond weakly to systematic changes in the input
autocorrelation (while leaving the mean input untouched), or for which
the input is only weakly autocorrelated, \prettyref{eq:cox_crr-3}
can be simplified. Specifically, when 
\begin{equation}
\begin{aligned}\int dt^{\pprime}C_{\phi}(t^{\prime},t^{\pprime}) & K_{xC_{\phi}}(t,t^{\prime},t^{\pprime})\\
\ll\int & dt^{\pprime}C_{\eta}(t^{\prime},t^{\pprime})K_{xm}(t,t^{\pprime}),
\end{aligned}
\label{eq:cox_approx_requirement}
\end{equation}
we find the useful relation
\begin{equation}
C_{x\eta}(t,t^{\prime})\approx\int dt^{\pprime}C_{\eta}(t^{\prime},t^{\pprime})K_{xm}(t,t^{\pprime})\label{eq:approx_color_crr}
\end{equation}
or, for stationary processes in frequency domain,
\begin{equation}
S_{x\eta}(\omega)\approx S_{\eta}(\omega)\chi_{m}(\omega).\label{eq:approx_color_crr_stationary}
\end{equation}
The approximate CRR for colored shot noise \prettyref{eq:approx_color_crr}
takes the same form as the Gaussian FNT \prettyref{eq:fnt_orig},
but note the differences between controlling the mean of Gaussian
noise and controlling the intensity of shot noise, as mentioned below
\prettyref{eq:fnt_orig}.

The approximation is exact for linear systems, for which always $K_{xC_{\phi}}\equiv0$.
In \prettyref{subsec:Leaky-integrate-and-fire-neuron} we discuss
a nonlinear system for which the approximation also holds (see \prettyref{fig:cox}),
and in \prettyref{subsec:Recurrent-neural-network} we show that \prettyref{eq:approx_color_crr}
is an improvement over \prettyref{eq:crr} even in a situation where
the colored shot-noise input is not a Cox process.

\section{Applications of the CRR\label{sec:Applications}}

\begin{figure}
\includegraphics{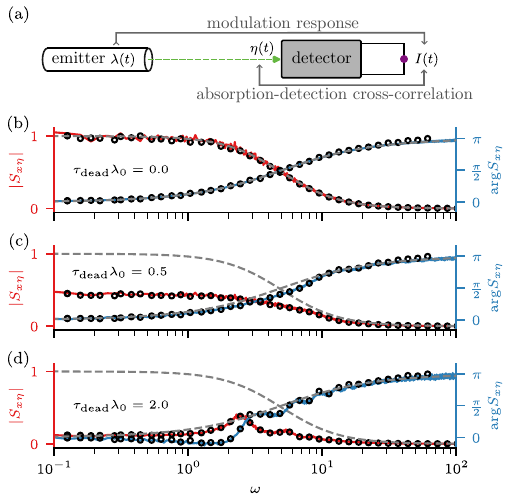}

\caption{\label{fig:detector}Test of the cross-correlation--response relation
\prettyref{eq:crr_stationary} for a particle detector without (b)
and with deadtime (c,d) as described in \prettyref{subsec:detector}
and sketched in (a). Both, the cross-spectrum $S_{I\eta}$ (red: absolute
value, blue: argument) and the susceptibility $\lambda_{0}\chi_{I\eta}$
(black circles) are determined from stochastic simulations (\prettyref{appsec:Numerical-methods})
and averaged over $1000$ trials -- their agreement corroborates
\prettyref{eq:crr_stationary}. (b-d) test this agreement for increasing
values of the detector deadtime, making the system increasingly nonlinear,
yet \prettyref{eq:crr_stationary} is respected in that colored lines
and black circles agree. The gray dashed lines show the exact cross-spectrum
for $\tau_{\text{dead}}=0$, that is the absolute value and argument
of $\lambda_{0}\protect\cF[\alpha](\omega)$, where $\protect\cF[\alpha](\omega)=(1-i\omega\tau_{p})^{-2}$
is the Fourier transform of \prettyref{eq:detector_pulse_alpha}.
Parameters: $\lambda_{0}=1$, $\tau_{p}=0.2$, $\tau_{\zeta}=1$,
$\sigma_{\zeta}^{2}=0.01$, $\Delta t=10^{-4}$, $T=200$, $T_{\text{warm}}=50$.}
\end{figure}

\subsection{Particle detector\label{subsec:detector}}

Detectors resolving single particles in a beam, like electrons or
photons, are of unquestionable importance: from the early Geiger M\"uller
counters \citep{turner2008atoms}, used to quantify radioactive decay,
to the newest superconducting-nanowire single-photon-detectors \citep{Natarajan2012_063001},
key elements for optical quantum computing, to name only two prominent
examples. These single particle detectors have the following issues
in common \citep{Stever1942_52,Valivarthi2014_24506}: (i) after detecting
one particle there is a non-vanishing deadtime within which subsequent
particle detection is unreliable and (ii) they are subject to thermal
noise, which is problematic at the low energies corresponding to single
particles. Here, we consider a generic detector model describing an
observable current 
\begin{equation}
I(t)=[\alpha\ast(\vartheta[\eta]\eta)](t)+\zeta(t)\label{eq:particle_detector}
\end{equation}
 which generates a pulse 
\begin{equation}
\alpha(t-t_{i})=\Theta(t-t_{i})\frac{t-t_{i}}{\tau_{p}^{2}}\exp\left[-\frac{t-t_{i}}{\tau_{p}}\right]\label{eq:detector_pulse_alpha}
\end{equation}
upon particle arrival at $t_{i}$, provided the detector is not in
a refractory state (dead) due to a preceding event. Specifically,
we consider a nonparalyzable detector \citep{turner2008atoms}, which
means that the switch $\vartheta$ attains the values $\vartheta[\eta;t]=0$
if $\int_{t-\tau_{\text{dead}}}^{t}\vartheta[\eta;t^{\prime}]\eta(t^{\prime})dt^{\prime}=1$
and $\vartheta[\eta;t]=1$ else. The initial value of the switch,
e.g. $\vartheta[\eta;0]=1$ corresponding to an initially susceptible
detector, is forgotten after $\approx\tau_{\text{dead}}(\tau_{\text{dead}}\lambda_{0})(1+\tau_{\text{dead}}\lambda_{0})$.
We also take into account thermal fluctuations in the form of an Ornstein-Uhlenbeck
process
\begin{equation}
\tau_{\zeta}\dot{\zeta}=-\zeta+\sqrt{2\sigma_{\zeta}^{2}\tau_{\zeta}}\xi(t)\label{eq:ou_process}
\end{equation}
{[}here, $\xi(t)$ is standard Gaussian white noise{]}.

Particle emission is often considered a Poisson process, matching
the key assumption in the derivation of the CRR. For instance, in
the case of coherent laser light, the photon absorption statistics
is exactly Poissonian \citep{MandelWolf1995optical}. Thus we may
use \prettyref{eq:crr_stationary} to relate the cross-spectrum $S_{I\eta}$
between the incoming particles and the elicited current to the susceptibility
$\chi_{I\eta}$ with respect to modulating the beam intensity $\lambda_{0}+\epsilon\nu(t)$,
namely $S_{I\eta}=\lambda_{0}\chi_{I\eta}$. We note that the independent
thermal noise $\zeta(t)$ does not affect the relation according to
what we discussed in \prettyref{subsec:Additional-noise}.

In \prettyref{fig:detector} we show the left- and right-hand sides
of \prettyref{eq:crr_stationary} for three different values of the
detector's deadtime, where the statistics are measured in stochastic
simulations, on which we elaborate in \prettyref{appsec:Numerical-methods}.
For $\tau_{\text{dead}}=0$, we have the simple example of linearly
low-pass filtered shot noise, for which both, the input-output cross-spectrum
and the susceptibility drop monotonically with frequency. The system
becomes nonlinear for nonvanishing deadtimes. As we see in \prettyref{fig:detector}(c)~and~(d)
the detector is less susceptible in this regime, i.e. we observe an
overall reduction of both the susceptibility and the cross-spectrum
and for long deadtimes {[}\prettyref{fig:detector}(d){]}, we even
see maxima forming in the spectral measures. In all cases, the CRR
is excellently confirmed by the simulation results. As sketched in
\prettyref{fig:detector}(a), in an experiment, the accessible manipulation
of the beam intensity and the observable mean response to it thus
permit to determine the exact cross-correlation between the actual
particle arrivals and the detector signal.

\begin{figure}
\includegraphics{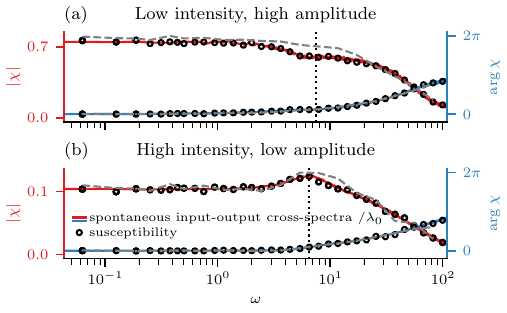}

\caption{\label{fig:CRR}LIF neuron model: Test of the cross-correlation--response
relation \prettyref{eq:crr_stationary} for a constant baseline intensity
$\lambda_{0}$ and the two cases of low intensity $\lambda_{0}$ and
high amplitude $A$ (a) and high intensity and low amplitude (b).
Both, the spontaneous input-output cross-spectra $S_{x\eta}/\lambda_{0}$
(red lines: absolute value, blue lines and right axis: argument),
and the susceptibility $\chi(\omega)$ (black circles, absolute value
and argument respectively) are determined from numerical simulations
(\prettyref{appsec:Numerical-methods}); for the susceptibility the
intensity is explicitly modulated ($\protect\epsilon>0$). The agreement
of colored lines and black circles corroborates \prettyref{eq:crr_stationary}.
Susceptibility based on the diffusion approximation \prettyref{eq:diff_approx_response}
(gray dashed line). Output firing rate (black dotted vertical lines).
\prettyref{eq:model} is integrated by Euler's method with time step
$\Delta t=10^{-4}$ and integration length $T=100$. Furthermore,
$v_{T}=1$, $v_{R}=0.5$, $\mu=0.5$, and $\tau_{s}=0.02$. In (a),
$\lambda_{0}=2$, $A=0.4$, in (b) $\lambda_{0}=16$, $A=0.05$. The
stimulus used to compute the susceptibility (circles) is $\protect\epsilon\cos(\omega_{s}t)$
with $\protect\epsilon=0.1$ (a) and $\protect\epsilon=0.16$ (b).}
\end{figure}

\subsection{Leaky integrate-and-fire neuron\label{subsec:Leaky-integrate-and-fire-neuron}}

Integrate-and-fire neurons are an abstraction of neural dynamics
that is excellent at predicting neural spike trains \citep{Brette2005_3642,Badel2008_370,Badel2008_666,Jolivet2008_426,Teeter2018_709}
and frequently used in simulations and in theory of spiking neural
networks \citep{AbbVre93,Bru00,SchDie15,Ost14,WieBer15,Ocker2023_041047_republ,Layer2024_013013}.
Such models exhibit a multitude of biologically observed network states
(see \citep{Bru00,Ost14} and references therein).

Here, we consider a leaky integrate-and-fire (LIF) neuron. In this
model, the neuron maps the Poissonian input spikes $\eta(t)=\sum_{i}\delta(t-t_{i})$
to its output by integrating the equation
\begin{equation}
\tau_{m}\dot{v}(t)=-v(t)+\mu+\left(\alpha\ast\eta\right)(t)\label{eq:model}
\end{equation}
until the membrane voltage $v(t)$ hits a threshold $v(t_{i}^{\text{IF}})=v_{T}$
at which the neuron spikes and is reset $v(t_{i}^{\text{IF}})\rightarrow v_{R}$,
see \prettyref{fig:models}(b). Time is counted in multiples of the
membrane time constant $\tau_{m}$, which to ease the notation, we
set to one. The neuron receives a constant current $\mu$, and the
input spikes are convolved with a synaptic filter $\alpha(t)=A\,\Theta(t)\,t\exp(-t/\tau_{s})/\tau_{s}^{2}$,
where $A$ is the synaptic amplitude and $\tau_{s}\ll1$ is the synaptic
timescale. The output spike train $x[\eta;t]=\sum_{i}\delta(t-t_{i}^{\text{IF}}),$
where $t_{i}^{\text{IF}}$ are the fire-and-reset times, is communicated
to other neurons and is thus the observable we are interested in.

\begin{figure}
\includegraphics{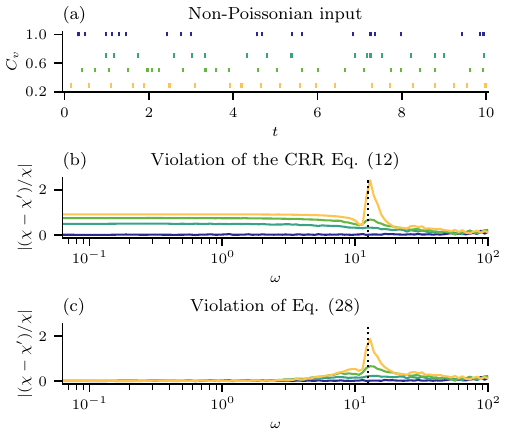}

\caption{\label{fig:Violation}LIF neuron model: Violation of the CRR \prettyref{eq:crr_stationary}
for non-Poissonian input. (a) Samples of a Poissonian process (dark
blue) and non-Poissonian processes ($C_{v}<1$). (b) Absolute value
of the relative error of \prettyref{eq:crr} $\left|\left[\chi(\omega)-S_{x\eta}(\omega)/\lambda_{0}\right]/\chi(\omega)\right|$
where $\chi(\omega)$ is the response to a modulation and $S_{x\eta}(\omega)/\lambda_{0}$
is the spontaneous cross-spectrum. As in \prettyref{fig:CRR}, both
the susceptibility $\chi$ and the input-output cross-spectrum $S_{x\eta}$
are obtained from numerical simulations, see \prettyref{appsec:Numerical-methods}.
For all but the lowest line, the input is non-Poisson with colors
corresponding to (a). The dotted line indicates the input intensity
$2\pi\lambda_{0}$. (c) Absolute value of the error $\left|\left[\chi(\omega)-S_{x\eta}(\omega)/S_{\eta}(\omega)\right]/\chi(\omega)\right|$,
which quantifies the mismatch of \prettyref{eq:approx_color_crr_stationary},
for the same setting as in (b). The vanishing error at small $\omega$
indicates that here \prettyref{eq:approx_color_crr_stationary} is
a successful color-correction of \prettyref{eq:crr} whereas the remaining
error about the input intensity $\omega=2\pi\lambda_{0}$ indicates
that here the naive color-correction fails. Parameters as in \prettyref{fig:CRR}(a).}
\end{figure}

\subsubsection{Test of the CRR and limitations for non-Poissonian input spikes}

In \prettyref{fig:CRR} we present simulation results for the cross-spectrum
$S_{x\eta}(\omega)$ using spontaneous input ($\epsilon=0$) and the
susceptibility $\chi(\omega)$ using modulated input ($\epsilon>0$).
Both simulations are done for two distinct parameter sets (a, b).
For once, we test the case of a low input intensity with high amplitude;
here two input spikes in short succession are sufficient to trigger
an output spike. The susceptibility in this case is rather high and
decays with increasing frequency. Secondly, we use a high input intensity
with low amplitude of the input spikes (b), such that the mean input
is the same as in (a) but we are close to a diffusion limit \citep{Holden1976Models,Tuckwell1989stochastic}.
In this setting we observe a much lower susceptibility (by a factor
of $\approx7$ for low frequencies), which is due to the reduced synaptic
amplitude $A$ (this is similar to the findings in \citep{Richardson10_178102}).
In addition, the high-intensity case features a peak around the firing
rate that is caused by the reduction in the effective noise level.
Most importantly in the context here, in both of these opposite cases,
\prettyref{eq:crr_stationary} is excellently confirmed, i.e., the
cross-correlation between input and output spikes in the spontaneous
case fully agrees with the linear response to an intensity modulation.

A common approximation of shot-noise driven systems, the diffusion
approximation \citep{gardiner1985handbook}, replaces
an inhomogeneous Poisson process $\eta(t)$ with intensity $\lambda(t)+\epsilon s(t)$
by white Gaussian noise $z(t)$ with matched first- and second-order
statistics, i.e.,
\begin{equation}
\eta(t)\approx z(t)\equiv\lambda(t)+\epsilon s(t)+\sqrt{\lambda(t)+\epsilon s(t)}\xi(t),\label{eq:diff_approx_definitiion}
\end{equation}
where $\xi(t)$ is white Gaussian noise with $\left\langle \xi(t)\right\rangle =0$
and $\left\langle \xi(t)\xi(t^{\prime})\right\rangle =\delta(t-t^{\prime})$.
This replacement only becomes exact in a very specific scaling limit
(see \prettyref{appsec:CRR-implies-white-noise}), but is often trusted
to be a good approximation. If one seeks to estimate the linear response
of $\left\langle x[\eta;t]\right\rangle $ to modulations of the intensity
of the Poisson process using the diffusion approximation, it becomes
apparent from \prettyref{eq:diff_approx_definitiion} that one has
to compute the linear response of $\left\langle x[z;t]\right\rangle $
to simultaneous modulation of both, the mean and the noise intensity
of the Gaussian process $z(t)$. We present this diffusion-approximation-based
linear response
\begin{equation}
K_{x\lambda}^{\text{da}}(t,t^{\prime})\equiv\frac{\delta}{\delta\lambda(t^{\prime})}\left[\left\langle x[z;t]\right\rangle _{z\sim\mathcal{N}[\lambda(t),\lambda(t)]}\right]\label{eq:diff_approx_response}
\end{equation}
alongside the true response to intensity modulations in \prettyref{fig:CRR}.
To this end, we simulate \prettyref{eq:model} but with the Gaussian
process above replacing $\eta$ and simultaneously modulate both the
mean and the noise intensity (results shown in \prettyref{fig:CRR}
as gray dashed lines). For the parameters that we have chosen, the
diffusion approximation turns out to be a reasonable description of
the susceptibility, although in the case of low input intensity and
high amplitudes {[}panel (a){]}, somewhat stronger deviations become
apparent.

\emph{}The validity of \prettyref{eq:crr} and thereby \prettyref{eq:crr_stationary}
relies on the Poissonianity of the input. To exemplify this, we
generate non-Poissonian input processes with intensity $\lambda_{0}+\epsilon\nu(t)$,
by first sampling a Poisson process with intensity $n\,\left[\lambda_{0}+\epsilon\nu(t)\right]$
and then keeping only every $n$'th spike. As shown in \prettyref{fig:Violation}(a),
this procedure, in the absence of modulation $\epsilon=0$, decreases
the coefficient of variation $C_{v}^{2}=\left\langle \left\langle I^{2}\right\rangle \right\rangle /\left\langle I\right\rangle ^{2}=1/n$
(where $\left\langle \left\langle \cdot\right\rangle \right\rangle $
denotes a cumulant, i.e., here the variance) of the interspike interval
$I$ such that the so generated processes (for $n=2,3,...$) are more
regular than a Poisson process ($C_{v}=1$). For these processes,
\prettyref{eq:crr_stationary} is violated, as shown in \prettyref{fig:Violation}(b).
Moreover, no frequency-independent linear relation between $S_{x\eta}$
and $\chi$ can be found. The violation is most severe in a frequency
band about the input intensity (black dotted line).
\begin{figure}
\includegraphics{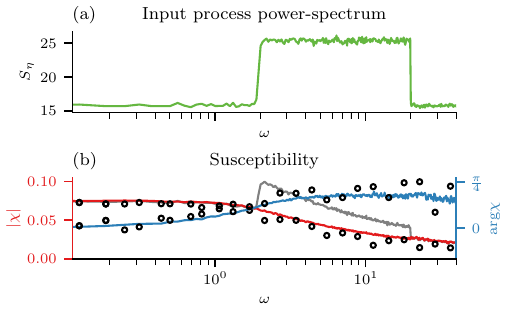}

\caption{\label{fig:cox}Test of the approximate CRR, \prettyref{eq:approx_color_crr_stationary},
for a Cox-process--driven LIF neuron. (a) Power spectrum $S_{\eta}$
of a Cox process, in which the intensity is band-pass Gaussian noise
with mean $m(t)=16$ and power spectrum $S_{\lambda}(\omega)=10\cdot\mathds{1}_{2<\omega<20}$.
(b) Absolute value (red) and argument (blue) of the right-hand side
of \prettyref{eq:approx_color_crr_stationary}, and respective susceptibility
of the model \prettyref{eq:model} (black circles), obtained by stochastic
simulations (\prettyref{appsec:Numerical-methods}) using the stimulus
$\protect\epsilon\cos(\omega_{s}t)$ with $\protect\epsilon=0.2$.
The gray line is the absolute value of the right-hand side of \prettyref{eq:crr_stationary}.
Parameters: Time step $\Delta t=10^{-3}$, integration time $T=100$,
$v_{T}=1$, $v_{R}=0.5$, $\mu=0$, $\tau_{s}=0$, $A=0.05$.}
\end{figure}

The deviation in \prettyref{fig:Violation}(b) can be cured at low
frequencies by using naively the color-correction \prettyref{eq:approx_color_crr_stationary},
but remains at frequencies $\omega\apprge2\pi\lambda_{0}$ {[}see
\prettyref{fig:Violation}(c){]}. However, if the input process is
truly a Cox process, as required in \prettyref{subsec:cox}, the color-correction
\prettyref{eq:approx_color_crr_stationary} seems to be a good approximation,
see \prettyref{fig:cox}. Thus, the correlation-response $K_{C}$
seems to be negligible for the model \prettyref{eq:model}. In \prettyref{subsec:Recurrent-neural-network}
we find a situation in which \prettyref{eq:approx_color_crr_stationary}
works for non-Cox noise, too.

\subsubsection{Fluctuation-response relation}

\begin{figure}
\includegraphics{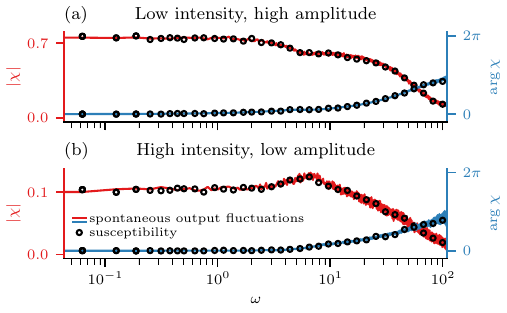}

\caption{\label{fig:FRR}Test of the fluctuation-response relation, \prettyref{eq:frr},
for a constant baseline intensity $\lambda_{0}$. (a) and (b) Susceptibility
computed from spontaneous output fluctuations {[}right-hand side of
\prettyref{eq:frr}{]}, absolute value (red) and argument (blue, right
axis), and susceptibility computed by explicitly modulating the intensity
(black circles, absolute value and argument respectively). The agreement
of colored lines and black circles corroborates \prettyref{eq:frr}.
The susceptibility computed by explicit modulations (black circles)
is the same as in \prettyref{fig:CRR}, but the colored lines here
are exclusively \emph{output} statistics. All statistics are measured
in stochastic simulations, see \prettyref{appsec:Numerical-methods}.
Parameters as in \prettyref{fig:CRR}, except for the spontaneous
case in (b), here $\Delta t=10^{-5}$ was necessary to achieve agreement.}
\end{figure}
Next, we leverage \prettyref{eq:crr_stationary} and follow the approach
of \citep{Lindner22_198101}, to derive a relation between the spontaneous
output fluctuations of the model \prettyref{eq:model} and the output's
response to modulations of the input intensity. To this end, we formally
incorporate the reset mechanism into \prettyref{eq:model}
\begin{equation}
\dot{v}=-v+\mu+(\alpha\ast\eta)-(v_{T}-v_{R})x(t).\label{eq:model_with_reset}
\end{equation}
Note that if the input would not be smoothed by $\alpha$, the reset
term would have to be $-[v(t)-v_{R})]x(t)$, as e.g. in Ref. \citep{Ocker2023_041047_republ},
to account for overshooting. The product $v(t)x(t)$ would be inconvenient
because it would lead to third-order statistics in the following expressions
(a similar problem emerges when an absolute refractory period is taken
into account \citep{Puttkammer2024_0770}), which we avoid by using
the synaptic filter, which is biophysically more plausible anyway.
Following \citep{Lindner22_198101}, we assume stationary statistics
(i.e., $\lambda(t)=\lambda_{0})$, apply Rice's method to \prettyref{eq:model_with_reset},
and get for $\omega\neq0$
\begin{equation}
S_{x\eta}(\omega)=\frac{(v_{T}-v_{R})S_{x}(\omega)+(1+i\omega)S_{xv}(\omega)}{\tilde{\alpha}^{\ast}(\omega)},\label{eq:relating_spectra}
\end{equation}
where $\tilde{\alpha}(\omega)=A\left(1-i\omega\tau_{s}\right)^{-2}$,
the (cross-) power spectra $S_{FG}(\omega)=\left\langle \left\langle \tilde{F}(\omega)\tilde{G}^{\ast}(\omega)\right\rangle \right\rangle /T$
with the finite-time-window Fourier transform $\tilde{G}(\omega)=\int_{0}^{T}dt\,e^{i\omega t}G(t)$,
and the asterisk denotes the complex conjugate.

Plugging \prettyref{eq:crr_stationary} into \prettyref{eq:relating_spectra}
then yields a fluctuation-response relation (FRR)
\begin{equation}
\chi(\omega)=\frac{(v_{T}-v_{R})S_{x}(\omega)+(1+i\omega)S_{xv}(\omega)}{\lambda_{0}\tilde{\alpha}^{\ast}(\omega)}.\label{eq:frr}
\end{equation}
Thus, the susceptibility can be computed using exclusively output
fluctuations, quantified by $S_{x}$ and $S_{xv}$, and without knowing
the input spike times (although $\lambda_{0}$ and $\alpha$ must
still be known). \prettyref{eq:frr} is tested and confirmed in \prettyref{fig:FRR}.
\begin{figure}
\includegraphics{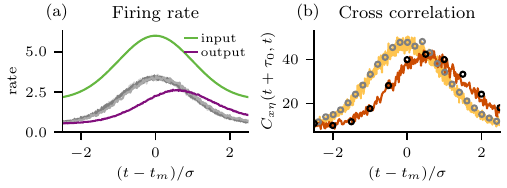}

\caption{\label{fig:Non-stationary-dynamics.}LIF neuron model: Test of the
\prettyref{eq:crr} for non-stationary dynamics. (a) Output firing
rate (gray and purple) of \prettyref{eq:model} for a time-dependent
input intensity $\lambda(t)=2+4\exp[-(t-t_{m})^{2}/(2\sigma^{2})]$
(green) for $\sigma=10$, $\sigma=5$, and $\sigma=1$ (overlapping
gray lines) and $\sigma=0.1$ (purple line). (b) Cross-correlation
$C_{x\eta}(t+\tau_{0},t)$ with $\tau_{0}=0.05$ for $\sigma=5$ (yellow
line) and $\sigma=0.1$ (orange line), and response function $\lambda(t)K(t+\tau_{0},t)$
for $\sigma=5$ (gray circles) and $\sigma=0.1$ (black circles).
Furthermore $v_{T}=1$, $v_{R}=0.5$, $\mu=0$, and $\tau_{s}=0.02$.}
\end{figure}

\subsubsection{A non-stationary case}

While many tools for stochastic systems are tailored to stationary
situations, in a number of areas, such as climate research or biology,
non-stationary behavior cannot be ignored without loosing key features
of the dynamics. For instance, the model \prettyref{eq:model} does
not approach stationarity if the baseline rate $\lambda(t)$ is not
constant. If $\lambda(t)$ varies sufficiently slowly, one would not
expect strong differences from a stationary setting. Indeed, as we
show in \prettyref{fig:Non-stationary-dynamics.}, when incorporating
a slow Gaussian pulse $\lambda(t)=\lambda_{01}+\lambda_{02}\exp[-(t-t_{m})^{2}/(2\sigma^{2})]$
into the baseline intensity, the system's output rate adapts adiabatically
to the changing input intensity. This is reflected by the scale invariance
of the output rate with respect to the pulse width $\sigma$, see
the overlapping gray lines in \prettyref{fig:Non-stationary-dynamics.}(a).
Thus, non-surprisingly, the non-stationary CRR \prettyref{eq:crr}
is fulfilled for the adiabatic case {[}yellow line and gray circles
agree in \prettyref{fig:Non-stationary-dynamics.}(b){]}. Truly interesting
non-stationary behavior is achieved, when the pulse is too short to
be responded to adiabatically, see the purple line in \prettyref{fig:Non-stationary-dynamics.}(a),
which breaks the scale invariance (and also the symmetry w.r.t. the
pulse center $t_{m}$). However, as we confirm for $\sigma=0.1$,
\prettyref{eq:crr} is still valid in this non-adiabatic case {[}orange
line and black circles agree in \prettyref{fig:Non-stationary-dynamics.}(b){]}.

\subsection{Recurrent neural network\label{subsec:Recurrent-neural-network}}

Here, we discuss the problem of stimulating a `control' neuron in
a network to achieve a desired time-dependent firing rate in a `target'
neuron. If the susceptibility of the control neuron $\chi_{cI}$ to
a current injection $I(t)$ is known and spontaneous measurements
of pair-wise cross-correlations have been conducted, one can apply
the approximate CRR for colored shot noise \prettyref{eq:approx_color_crr_stationary}
to estimate the remote susceptibility 
\begin{equation}
\chi_{tI}(\omega)\approx S_{c}(\omega)^{-1}S_{tc}(\omega)\chi_{cI}(\omega),\label{eq:remote_response_Poiss}
\end{equation}
where $S_{c}$ is the power spectrum of the control neuron and $S_{tc}$
is the cross-spectrum between the target- and the control neuron.
Knowledge of $\chi_{tI}$ then allows one to make the target fire
with a desired rate $r_{t}(t)$ by applying the current 
\begin{equation}
I(t)=\cF^{-1}\left[\cF[r_{t}]/\chi_{tI}\right](t)\label{eq:compute_current}
\end{equation}
to the control neuron. Note that beside the assumptions for \prettyref{eq:approx_color_crr_stationary},
one also needs to assume that the control and the target neuron receive
independent noise from the rest of the network (see \prettyref{subsec:Additional-noise}).
This can be violated in dense networks, but, as we show next, for
the biologically relevant case of sparse neural networks the assumption
is justified.

For concreteness, we consider a sparsely connected random neural network
\citep{Bru00}. This network model consists of $N_{E}$ excitatory
and $N_{I}=N_{E}/4$ inhibitory LIF neurons. Each neuron has exactly
$C_{E}$ incoming excitatory synapses with efficacy $J$, and $C_{I}$
incoming inhibitory synapses with efficacy $-gJ$. Thus the evolution
of the network is given by 
\begin{equation}
\begin{aligned}\dot{v}_{i}=-v_{i}+J\sum_{j\in C_{E}(i)}x_{j}(t)- & gJ\sum_{j\in C_{I}(i)}x_{j}(t)\\
+J\sum_{j=1}^{C_{E}}x_{\text{ext},j}^{i},
\end{aligned}
\label{eq:brunel_net}
\end{equation}
with the additional fire-and-reset rule as in \prettyref{eq:model}.
Here, $x_{\text{ext},j}^{i}$ are independent external Poisson processes
with intensity $\nu_{\text{ext}}$ and $C_{E}(i)$ {[}$C_{I}(i)${]}
is the set of excitatory {[}inhibitory{]} neurons that send spikes
to neuron $i$.

\begin{figure}
\includegraphics{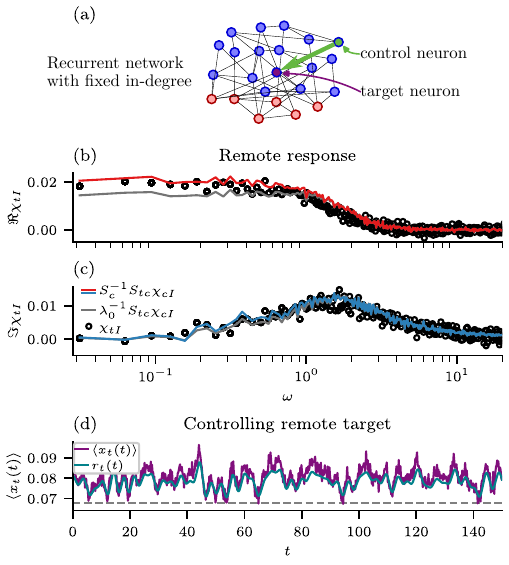}

\caption{\label{fig:Remote-control}Remote control in recurrent neural networks.
(a) Network model \prettyref{eq:brunel_net}. One random neuron is
selected as `control' neuron, another random neuron is selected as
`target' neuron, constrained to be at one synapse distance from control
(green arrow). The control neuron receives the current $I(t)$ in
\prettyref{eq:compute_current} to make the target fire with a desired
rate $r_{t}(t)$. (b) and (c) Real- and imaginary part of the target's
response to control stimulation measured by stimulation (black circles)
and estimated from \prettyref{eq:remote_response_Poiss} (red and
blue lines). The gray line shows the estimate based on \prettyref{eq:crr_stationary},
i.e. without the color correction \prettyref{eq:approx_color_crr_stationary}.
(d) Desired rate $r_{t}(t)$ (turquoise line) and achieved rate (purple
line) of the target neuron after application of the current \prettyref{eq:compute_current}
to the control neuron. The spontaneous rate of the target neuron is
$r_{0}\approx0.68$ (gray dashed line). Parameters: $N_{E}=10^{5}$,
$N_{I}=2.5\cdot10^{4}$, $C_{E}=200$, $C_{I}=50$, $g=4.2$, $J=0.01$,
$\nu_{\text{ext}}=0.83/(JC_{E})$. The spectral measures in (b) and
(c) are averaged over $10^{4}$ noise realizations and the rate in
(d) is averaged over $10^{3}$ noise realizations. In both cases,
the network realization is fixed. For a membrane time-constant of
$\tau_{m}=\protect\Qty{10}{\milli\second}$, the y-ticks in (d) are
$\protect\Qty 7{\hertz}$, $\protect\Qty 8{\hertz}$, and $\protect\Qty 9{\hertz}$
respectively.}
\end{figure}

In \prettyref{fig:Remote-control}, we show simulation results for
this network. We selected two neurons as control and target respectively,
enforcing that the target is at one-synapse distance from the control
{[}highlighted by the green arrow in \prettyref{fig:Remote-control}(a){]}.
The susceptibilities $\chi_{cI}$ and $\chi_{tI}$ are obtained from
simulations in which the control neuron was directly stimulated with
a Gaussian white noise current. The cross-spectrum $S_{tc}$ and the
power spectrum $S_{c}$ are obtained from spontaneous simulations.
As shown in \prettyref{fig:Remote-control}(b)~and~(c), the estimate
\prettyref{eq:remote_response_Poiss} captures the susceptibility
quite well. In \prettyref{fig:Remote-control}(d), we then demonstrate
how $\chi_{tI}$ can be exploited to control the target indirectly:
Here, we generate a Gaussian broadband stimulus $r_{t}(t)$ (turquoise
line) with cut-off frequency $\omega_{c}=2$, feed the current \prettyref{eq:compute_current}
to the control neuron, and observe the firing rate (purple line) of
the target neuron. Note that the parameters here are chosen such
that for a reasonable membrane time constant $\tau_{m}=\Qty{10}{\milli\second}$,
the spontaneous rate in dimensional units is $\approx\Qty 7{\hertz}$,
which is biologically reasonable.

\section{Further variants of the CRR}

Here we discuss two further variants of the CRR: First we show
how to include random amplitudes of the input spikes, then we show
CRRs for higher-order input-output cross-correlation functions and
nonlinear response functions.  Given the characteristic functional
$Z_{\boldsymbol{p}}[u]$ of the input process, where $\boldsymbol{p}$
denotes the (time-dependent) parameters of the noise model, CRRs can
be found systematically by recognizing that they correspond to relations
between functional derivatives of $Z_{\boldsymbol{p}}[u]$ w.r.t.
$u(t)$ and functional Taylor coefficients of $Z_{\boldsymbol{p}}[u]$
w.r.t. parameters $p_{i}$. This is analogous to the derivations of
\prettyref{eq:crr} and \prettyref{eq:cox_crr-3} and will be exemplified
for the two further cases below.

\subsection{Random amplitudes of input spikes\label{subsec:Random-amplitudes}}

As a first variant, we consider an input process with random amplitudes
$\eta(t)=\sum_{i}a_{i}\delta(t-t_{i})$ where $t_{i}$ are Poisson
events with intensity $\lambda_{0}$ and $a_{i}$ are independently
sampled from an exponential distribution $p(a)=\Theta(a)b^{-1}\exp(-a/b)$.
Thus, we replace the Poissonian input with a \emph{marked} Poisson
process. In the example of the shot-noise--driven LIF neuron, random
amplitudes are a more faithful description of synaptic inputs that
display considerable variability \citep{Tuckwell1983_110,Koc99,Manwani1999_1829,Richardson10_178102}.
Note that while one could ascribe the random amplitudes to an additional
random process as described in \prettyref{subsec:Additional-noise},
specifically $\eta(t)=\xi(t)\eta_{0}(t)$, where $\eta_{0}$ is the
unmarked Poisson process and $\xi(t)\overset{\text{i.i.d.}}{\sim}\Theta[\xi(t)]b^{-1}\exp[-\xi(t)/b]$,
where independence refers to the time argument, the CRR \prettyref{eq:crr}
only applies to relations between the output and $\eta_{0}$, whereas
here we study the relation between the output and the full marked
input $\eta(t)$.

The mean of the input process is $\left\langle \eta(t)\right\rangle =b\lambda_{0}$,
thus modulating $\lambda_{0}$ and $b$ has a similar effect on the
mean input. In \citep{Richardson10_178102}, the susceptibility $\chi(\omega)$
of a LIF neuron to an intensity modulation of such an input process
has been derived. To fully explain the input-output cross-spectrum
$S_{x\eta}(\omega)$ it turns out that $\chi(\omega)$ is not sufficient.
Specifically, we show that one also needs the susceptibility $\chi_{b}(\omega)$
to time-dependent modulations of $b$.

Our starting point is the characteristic functional of an independently
and identically marked Poisson process \citep{snyder2012}
\begin{equation}
Z_{\lambda,b}[u]=e^{\int\lambda(t^{\prime})\left[\phi_{b}\left(u(t^{\prime})\right)-1\right]dt^{\prime}},
\end{equation}
where $\phi_{b}(u)=\left\langle e^{iau}\right\rangle $ is the characteristic
function of the marks (here amplitudes). For exponentially distributed
amplitudes, $\phi_{b}(u)=1/(1-iub)$. Similarly to \prettyref{eq:cross_corr_from_char_func},
$C_{x\eta}$ can be expressed by a functional derivative $D(t)\equiv\frac{\delta}{\delta iu(t)}Z_{\lambda,b}[u]$,
and similarly to \prettyref{eq:lin_resp_by_char}, the two linear-response
functions can be expressed by functional Taylor coefficients $T(t)\equiv\left.\frac{\delta}{\delta\Lambda(t)}Z_{\Lambda,b}[u]\right|_{\Lambda=\lambda}$
and $T_{b}(t)\equiv\left.\frac{\delta}{\delta B(t)}Z_{\lambda,B}[u]\right|_{B=b}$.
Due to the identity (easily checked by insertion)
\begin{equation}
\frac{\partial}{\partial iu}\phi_{b}(u)=b\phi_{b}(u)+b^{2}\frac{\partial}{\partial b}\phi_{b}(u),
\end{equation}
the three functions $D$, $T$, and $T_{b}$ are directly related
\begin{equation}
D(t^{\prime})=\lambda bT(t^{\prime})+b^{2}T_{b}(t^{\prime}),\label{eq:helping_towards_rich_crr}
\end{equation}
as follows by straight forward differentiation. If we integrate \prettyref{eq:helping_towards_rich_crr}
with $\int\cD u\,y[u,t]\times$ and assume stationarity, we find for
$\omega\neq0$ the CRR
\begin{equation}
S_{x\eta}(\omega)=\lambda_{0}b\chi(\omega)+b^{2}\chi_{b}(\omega).\label{eq:richardson_crr}
\end{equation}
Thus, in the case of random amplitudes,the spontaneous input-output
cross-spectrum is connected to \emph{two} mechanistic properties of
the system, the linear responses to intensity- and amplitude modulations,
respectively.

\begin{figure}
\includegraphics{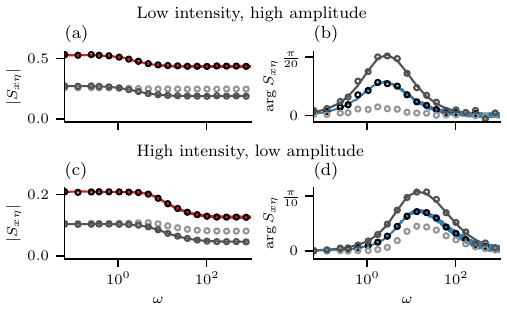}

\caption{\label{fig:rich_crr}LIF neuron model: Test of the CRR for a LIF neuron
driven by a Poisson process input with random amplitudes \prettyref{eq:richardson_crr}.
Absolute value {[}(a) and (c), red lines{]} and argument {[}(b) and
(d), blue lines{]} of the cross-spectrum $S_{x\eta}(\omega)$. Right-hand
side of \prettyref{eq:richardson_crr} (black circles) and single
contributions $b^{2}\chi_{b}(\omega)$ (light gray circles) and $\lambda_{0}b\chi(\omega)$
(dark gray circles); for the latter, the exact result \citep{Richardson10_178102}
is shown (dark gray line), for all other lines and symbols, the statistics
are measured in stochastic simulations, see \prettyref{appsec:Numerical-methods}.
Parameters: $v_{T}=1$, $v_{R}=0.5$, $\mu=0$, and $\tau_{s}=0$.
For (a) and (b) $\lambda_{0}=2$, $b=0.4$; and for (c) $\lambda_{0}=24$,
$b=1/20$.}
\end{figure}

The CRR, \prettyref{eq:richardson_crr}, is verified and illustrated
in \prettyref{fig:rich_crr} for two opposite cases of the input process.
The cross-spectrum decreases monotonically with frequency and saturates
at a non-vanishing level, corresponding to the event that an input
spike triggers immediately an output spike; this is more likely for
larger amplitudes {[}saturation is larger in (a) than in (c){]} and
relies on our choice of a vanishing filter time $\tau_{s}=0$. We
also illustrate in \prettyref{fig:rich_crr} the relevance of the
two contributions in \prettyref{eq:richardson_crr}, which agree for
small frequencies but deviate otherwise, especially pronounced in
the phase for large amplitudes and intermediate frequencies (b).

\subsection{Higher-order statistics and nonlinear response}

Lastly, we demonstrate a higher-order CRR involving nonlinear response
functions and higher-order cross-correlations. It has been recently
suggested \citep{Schlungbaum2023_108} that the nonlinear response
of sensory cells may be important in certain detection problems \citep{Henninger2018_5465}.
Furthermore, taking into account nonlinear response overcomes the
limitations of linear response theory of neural activity \citep{Ocker2017_47}
and may exhibit surprising features already for simple neuron models
\citep{VorLin17}.

Considering again a fixed-amplitude inhomogeneous Poisson process,
the third input-output cumulant $C_{x\eta\eta}(t,t^{\prime},t^{\pprime})=\left\langle \left\langle x(t)\eta(t^{\prime})\eta(t^{\prime\prime})\right\rangle \right\rangle $
can be attributed to a linear combination of the linear-response function
$K(t,t^{\prime})$ and the second order response function $K_{2}(t,t^{\prime},t^{\pprime})=\left.\frac{\delta^{2}}{\delta\epsilon s(t^{\prime})\delta\epsilon s(t^{\pprime})}\left\langle x[\eta;t]\right\rangle _{\epsilon}\right|_{\epsilon=0}$.
Analogously to the above derivation, expressing $C_{x\eta\eta}$,
$K_{2}$, and $K$ as path integrals including the second functional
derivative of the characteristic functional of $\eta$ and its second-
and first functional Taylor coefficients w.r.t. the intensity, respectively,
we find by comparing the integrands
\begin{equation}
\begin{aligned}C_{x\eta\eta}(t,t^{\prime},t^{\pprime}) & =\delta(t^{\prime}-t^{\prime\prime})\lambda(t^{\prime})K(t,t^{\prime})\\
 & \;+\lambda(t^{\prime})\lambda(t^{\pprime})K_{2}(t,t^{\prime},t^{\prime\prime}).
\end{aligned}
\end{equation}
This strikingly simple relation reveals that the third-order cross-correlations
are entirely determined by first- and second-order response functions.
Similarly, cross-correlations of order $n$ are given in terms of
response functions up to order $n-1$. Likewise, cross-correlations
including orders $m\geq2$ of $x$ are related to response functions
of $\prod_{i=1}^{m}x(t_{i})$.

\section{Summary and outlook}

\emph{}In this paper, we derived a number of exact relations between
the input-output cross-correlations of a shot-noise--driven system
and its response functions. These CRRs can be regarded as analogues
of the famous Furutsu-Novikov theorem for systems driven by Gaussian
noise but are, as we demonstrated, not the same. Our theorem, holding
true for the case of Poissonian shot noise (and in an extension for
Cox noise), applies to simple functionals such as a linear filter
but also to a more complicated system such as an excitable neuron
that itself generates spikes (i.e. another shot noise process). We
tested the basic relation for a particle detector and for a shot-noise--driven
integrate-and-fire neuron and demonstrated that it is nontrivial,
as it is not obeyed if the input shot noise deviates from Poisson
statistics. We used the CRR for this model class to derive a novel
FRR in the presence of shot noise. In a recurrent network we used
the CRR to extract remote-response functions from spontaneous cross-correlations.
Finally, we generalized the relation in two further respects: i) we
replaced the common Poissonian input noise by a marked Poisson process,
for which amplitudes are drawn from an exponential distribution; and
ii) we exemplified how higher-order cross-correlation functions of
shot-noise-driven systems can be related to higher-order response
functions.

The approaches developed here enable the derivation of families of
non-trivial input-output relations of systems driven by random series
of events. It is conceivable, for instance, that the generalizations
outlined above may be combined, i.e. we could consider a marked\textit{
and }doubly stochastic process as an input and also derive higher-order
relations in this setting. Another extension is the common situation
that a system is subject to several independent shot-noise processes
or to both shot noise and Gaussian noise. In neurons, for instance,
there are excitatory and inhibitory synaptic inputs and, moreover,
several competing types of noise, some of which can be approximated
by Gaussian noise, e.g. the channel noise from a large population
of independent ionic channels \citep{FoxLu94}. We expect that in
such cases, families of relations between various input-output cross-correlations
and response functions to various modulations (e.g. intensities of
excitatory and inhibitory input spike trains; mean and variance of
Gaussian input noise) can be found and may serve to derive, for instance,
corresponding families of FRRs. Another interesting model class for
neural activity is the Hawkes process \citep{Pernice2011_14,Jovanovic2016_28,Ocker2017_47},
which our study does not cover due to its deviation from Poissonian
input statistics. It is thus an open problem to derive CRRs for this
situation.

Regarding the specific application of the CRR to the integrate-and-fire
model, several remarks are in order. First of all, analytical results
for this model class are scarce, and the CRRs may allow us to derive
new exact results. For the case of random amplitudes, the intensity
response is known, see \citep{Richardson10_178102}, and the response
to modulations of the amplitude might be obtained by the methods therein;
knowing both of these functions would provide us with an explicit
expression for the input-output cross-correlation function of this
model. Secondly, the CRR can also be applied to more involved nonlinear
models, such as integrate-and-fire models with adaptation \citep{BreGer05,Izh03,ShiSch15,RamLin21},
with synaptic short-term plasticity (see e.g. \citep{AbbReg04,Lis97,MonBar08,LinGan09,RosRub12}),
or with conductance-based input shot noise \citep{RicGer05,RicGer06,WolLin08,LinLon06,Richardson2024_024407}.
Beyond the integrate-and-fire framework, CRRs may be exploited in
detailed biophysical models such as Hodgkin-Huxley--type neuron models
with a true spike-generating mechanism, and spatially extended neuron
models based on cable theory with stochastic inputs distributed over
the neuron's dendrite \citep{GowTim20,GowRic23}. Thirdly, at the
network level, input-output cross-correlations are particularly relevant,
because the input spikes for one neuron are another neuron's output
spikes, and modern multi-electrode arrays allow for parallel recording
of hundreds to thousands of spike trains. Expanding on the approach
worked out in \prettyref{subsec:Recurrent-neural-network}, we may
use the CRR to determine an entire matrix of pair-wise response functions
in heterogeneous networks of spiking neurons. Furthermore, from a
more theoretical point of view, CRRs can be helpful by constraining
the constituents in the theory of neural networks. For example, in
a recent cavity-method approach to rate-based neural networks \citep{Clark_23_118401}
with Gaussian statistics, the Gaussian FNT was used to connect neural
cross-correlations and response functions. With the results presented
here, such approaches can likely be extended to recurrent networks
of spiking neurons.\emph{ }Last but not least, in the theory of neural
learning, the important paradigm of spike-timing dependent plasticity
involves the cross-correlation of pre- and post-synaptic spike trains.
Relations such as the CRR constrain the possible dynamics of the synaptic
weights during learning and may thus be instrumental to understand
this type of self-organization in the brain.\emph{}
\begin{acknowledgments}
We are grateful to Igor Sokolov for helpful discussions. This work
has been funded by the Deutsche Forschungsgemeinschaft (DFG, German
Research Foundation), SFB1315, project-ID 327654276 to BL.
\end{acknowledgments}

\appendix

\section{\label{appsec:Numerical-methods}Numerical methods}

The main results of this manuscript are links between correlation
functions and response functions, both of which are in general hard
to obtain analytically. Therefore, in Figs.~2--9, we compute these
statistics from stochastic simulations. This involves sampling a large
number of realizations of the input processes, computing the respective
output processes, and assuming that averages over the ensemble of
input processes may be approximated by averages over the finite number
of input realizations. In this appendix, we lay out technical details.

\subsection{Generation of Poisson processes}

To sample Poisson processes with intensity $\Lambda(t)$ we discretize
time into timesteps $\Delta t$ and perform a Bernoulli trial in each
time bin $k$ with $t_{k}=k\Delta t$ \citep{snyder2012}. Specifically,
we draw uniform random numbers $q_{k}\sim\mathfrak{U}(0,1)$ between
$0$ and $1$. If $q_{k}\leq\Lambda(t_{k})\Delta t$, we note an event
at time $t_{k}$, thus the so-sampled Poisson process may be represented
by the discrete-time spike-train
\begin{equation}
\eta_{k}=\sum_{k^{\prime}\,\text{s.t.\,\ensuremath{q_{k^{\prime}}\leq\Lambda}(\ensuremath{t_{k^{\prime}}})\ensuremath{\Delta}t}}\frac{\delta_{kk^{\prime}}}{\Delta t}.
\end{equation}
Here, the ratio of the Kronecker delta and the time step is the discrete-time
approximation of the Dirac delta function in the continuous-time spike-train,
$\eta(t)=\sum\delta(t-t_{k^{\prime}})$.

\subsection{Integration of Langevin equations}

To sample the Ornstein-Uhlenbeck process \prettyref{eq:ou_process},
we integrate the Langevin equation using the Euler-Maruyama method.
Specifically, we initialize the value of the process at time $t_{0}$
as $\zeta_{0}=0$ and then iteratively compute the subsequent values
of the process
\begin{equation}
\zeta_{k+1}=\left(1-\frac{\Delta t}{\tau_{\zeta}}\right)\zeta_{k}+\sqrt{2\sigma_{\zeta}^{2}\frac{\Delta t}{\tau_{\zeta}}}z_{k},\label{eq:ou_process_discrete}
\end{equation}
where $z_{k}\overset{\text{i.i.d.}}{\sim}\mathcal{N}(0,1)$. To get
rid of transient behavior due to the choice of $\zeta_{0}$, we cut
off a transient time $T_{\text{warm}}=50\gg\tau_{\zeta}=1$. 

The same type of integration is applied to simulate the LIF dynamics
\prettyref{eq:model} driven by white Gaussian noise $z(t)$ \prettyref{eq:diff_approx_definitiion}
in order to determine the diffusion approximation of the response
function \prettyref{eq:diff_approx_response}. Here, we apply a Markovian
embedding to implement the filter $\alpha(t)$ by means of two additional
auxiliary variables $m_{1}$ and $m_{2}$
\begin{align}
\begin{aligned}(\alpha\ast\eta)(t)\approx(\alpha\ast z)(t) & =Am_{1}(t)/\tau_{s}^{2}\\
\dot{m}_{1}(t)= & -\tau_{s}^{-1}m_{1}(t)+m_{2}(t)\\
\dot{m}_{2}(t)= & -\tau_{s}^{-1}m_{2}(t)+z(t).
\end{aligned}
\label{eq:alpha_mark_emb}
\end{align}
The last line describes again an Ornstein-Uhlenbeck process that can
in discrete time be simulated as in \prettyref{eq:ou_process_discrete}.

\subsection{Sampling colored Gaussian noise}

For the production of \prettyref{fig:cox}, we use a different method
\citep[and references therein]{Dummer14} to sample the Gaussian input
process. Namely, we generate the Gaussian noise $\lambda(t)$ with
given mean $m$ and power spectrum $S_{\lambda}(\omega)$ by sampling
the Fourier transform $\tilde{\lambda}(\omega)$. Since we here consider
a stationary random process, that is $\left\langle \left\langle \lambda(t+\tau)\lambda(t)\right\rangle \right\rangle $
is invariant to translations in time $t$, the correlations in frequency
space are diagonal $\left\langle \left\langle \tilde{\lambda}(\omega)\tilde{\lambda}^{\ast}(\omega^{\prime})\right\rangle \right\rangle \propto\delta(\omega-\omega^{\prime})$.
Thus, for a finite time window $T$ and discrete frequency bins $\omega_{k}=2\pi k/T$,
we may conveniently sample $|\tilde{\lambda}(\omega_{k})|\sim\mathcal{N}\left[\tilde{m}(\omega_{k}),S_{\lambda}(\omega_{k})\right]$
and $\text{arg}[\tilde{\lambda}(\omega_{k})]\sim\mathfrak{U}(0,2\pi)$
independently for each frequency bin. Lastly we apply the inverse
Fourier transform to obtain $\lambda(t)$.

\subsection{Integrating LIF neurons}

To integrate the LIF neuron \prettyref{eq:model}, we first sample
the input as detailed above, and then integrate \prettyref{eq:model}
with a simple forward scheme. Specifically, at each step $k$, the
membrane voltage is updated as 
\begin{equation}
v_{k+1}=v_{k}+\Delta t\left[-v_{k}+\mu+\left(\alpha\ast\eta\right)(t_{k})\right]
\end{equation}
and the convolution $\alpha\ast\eta$ is computed in advance. Alternatively,
the Markovian embedding \prettyref{eq:alpha_mark_emb} could be used
for shot noise $\eta(t)$ instead of $z(t)$, too.

\subsection{Spontanous statistics}

For spontaneous statistics, e.g. the left-hand sides in Eqs. \eqref{eq:crr},
\eqref{eq:cox_crr-3}, \eqref{eq:frr}, and \eqref{eq:richardson_crr},
the simulations are computed as outlined above for large numbers of
input realizations. Per realization, the correlation functions and
spectra correspond to simple products (of Fourier transforms) of variables.
The presented statistics are averages of these per-realization products
over all realizations.

\subsection{Response functions}

The susceptibility $\chi$ of a variable $\left\langle x\right\rangle $
to modulations of a parameter $\lambda_{0}$ is computed as follows.
For each frequency $\omega_{s}$ at which the susceptibility is seeked,
we simulate the dynamics of $x$ for many realizations of the input
with parameter $\lambda_{0}+\epsilon\cos(\omega_{s}t)$. Assuming
stationary dynamics for $\epsilon=0$, linear response theory states
that for $\omega_{s}>0$
\begin{equation}
\left\langle \tilde{x}(\omega)\right\rangle =\epsilon\pi\ensuremath{\chi}(\ensuremath{\omega})\delta(\omega-\omega_{s})+\cO(\epsilon^{2}).
\end{equation}
Thus, for each $\omega_{s}$ we may extract the susceptibility $\chi(\omega_{s})=[\Delta\omega/(\epsilon\pi)]\left\langle \tilde{x}(\omega_{s})\right\rangle $,
where the average is taken over the realizations and the frequency
step is given in terms of the simulation window $T$ as $\Delta\omega=2\pi/T$.

\section{\label{appsec:CRR-implies-white-noise}CRR implies white-noise FNT
in the diffusion limit}

Here, we show that the CRR \prettyref{eq:crr} implies the FNT for
white Gaussian noise. To this end, we follow an approach put forward
by \citep{Lansky1984_647,Lansky1997_2043} to recall that every white
Gaussian noise process can be regarded as the diffusion limit of a
scaled Poisson process with offset. Thus, the CRR which applies to
Poisson processes has a corresponding property which applies to white
Gaussian noise; this property turns out to be the FNT.

The diffusion approximation in its common use refers to replacing
an inhomogeneous Poisson process by white Gaussian noise with matched
time-dependent mean and noise intensity. This replacement neglects
all cumulants of order $k\geq3$ of the Poisson process. However,
the scaled Poisson process with offset
\begin{equation}
z_{n}(t)=a_{n}\eta_{n}(t)+\psi_{n}(t),\label{eq:scaled_Poiss_with_offset}
\end{equation}
where $\eta_{n}(t)$ is a Poisson process with intensity $\lambda_{n}(t)$,
becomes equivalent to white Gaussian noise 
\begin{equation}
x[z_{n};t]\overset{n\rightarrow\infty}{\rightarrow}x[\xi;t],
\end{equation}
in the sense specified in \citep{Lansky1984_647,Lansky1997_2043},
in a specific limit. If 
\begin{equation}
\begin{aligned}\lambda_{n}(t) & =A(t)n+B(t)n^{2},\\
a_{n} & =a_{\dagger}/n,\\
\psi_{n}(t) & =-a_{\dagger}B(t)n,
\end{aligned}
\end{equation}
with $A$, $B$, and $a_{0}$ independent of $n$, then in the limit
$n\rightarrow\infty$ the mean and autocorrelation of $z$ are finite,
namely $\left\langle z(t)\right\rangle =a_{\dagger}A(t)$ and $\left\langle \left\langle z(t)z(t^{\prime})\right\rangle \right\rangle =a_{\dagger}^{2}B(t)\delta(t-t^{\prime})$,
yet all cumulant functions of order $k\geq3$ vanish $\propto1/\cO(n^{k-2})$.
Thus, in this limit, \prettyref{eq:scaled_Poiss_with_offset} is statistically
identical to a white Gaussian process $\xi$ with mean $m(t)=a_{\dagger}A(t)$
and noise intensity $D(t)=a_{\dagger}^{2}B(t)$. Conversely, every
white Gaussian noise process can be represented as the diffusion limit
of a scaled Poisson process with offset.

Modulating the intensity of the Poisson process, specifically replacing
$\lambda_{n}(t)$ by $\lambda_{n}(t)+\epsilon s_{n}(t)$ with $s_{n}(t)=s_{\dagger}(t)n$,
corresponds in the limit $n\rightarrow\infty$ to modulating the mean
of the Gaussian process, i.e. replacing $m(t)$ by $m(t)+\epsilon\hat{s}(t)$
with $\hat{s}(t)=a_{\dagger}s_{\dagger}(t)$, whereas the noise intensity
is not affected. Thus, the linear response $K_{n}$ of a functional
$x[\eta_{n};t]\equiv\hat{x}[z_{n}=a_{n}\eta_{n}+\psi_{n};t]$ to
modulation of $\lambda_{n}(t)$,
\begin{equation}
\begin{aligned}\left\langle x[\eta_{n};t]\right\rangle _{\epsilon}= & \left\langle x[\eta_{n};t]\right\rangle _{0}\\
 & +\epsilon\int dt^{\prime}K_{n}(t,t^{\prime})s_{n}(t^{\prime})+\cO(\epsilon^{2}),
\end{aligned}
\end{equation}
corresponds in the diffusion limit to the response of $\hat{x}$ to
modulating $m(t)$
\begin{align}
\left\langle \hat{x}[\xi;t]\right\rangle _{\epsilon}=\lim_{n\rightarrow\infty} & \left\langle \hat{x}[z_{n};t]\right\rangle _{\epsilon}\\
=\left\langle \hat{x}[\xi;t]\right\rangle _{0} & +\lim_{n\rightarrow\infty}\frac{\epsilon}{a_{n}}\int dt^{\prime}K_{n}(t,t^{\prime})\hat{s}(t^{\prime})+\cO(\epsilon^{2}),\nonumber 
\end{align}
where $\xi$ is the Gaussian noise introduced above. Thus the linear
response function to mean modulations of the Gaussian process is $K_{x\xi}=\lim_{n\rightarrow\infty}K_{n}/a_{n}$
in the diffusion limit. Additionally, the cross-correlations $C_{x\eta}(t,t^{\prime})=\left\langle \left\langle x(t)\eta_{n}(t^{\prime})\right\rangle \right\rangle $
and $C_{\hat{x}\xi}(t,t^{\prime})=\left\langle \left\langle \hat{x}(t)\xi(t^{\prime})\right\rangle \right\rangle $
are related by $C_{\hat{x}\xi}=a_{n}C_{x\eta}$ in the diffusion limit.
Thus, \prettyref{eq:crr} implies the white-noise case of the Gaussian
FNT \prettyref{eq:fnt_orig}
\begin{equation}
\begin{aligned}C_{\hat{x}\xi}(t,t^{\prime}) & =a_{n}^{2}\lambda_{n}(t^{\prime})K_{x\xi}(t,t^{\prime})\\
 & =D(t^{\prime})K_{x\xi}(t,t^{\prime}),
\end{aligned}
\label{eq:fnt_whitenoise}
\end{equation}
since $a_{n}^{2}\lambda_{n}(t)=\frac{a_{\dagger}^{2}}{n^{2}}(A(t)n+B(t)n^{2})\overset{n\rightarrow\infty}{\rightarrow}a_{\dagger}^{2}B(t)\equiv D(t)$.

\section{\label{appsec:Non-negative-cox}Impact of non-negativity}

\begin{figure}
\includegraphics{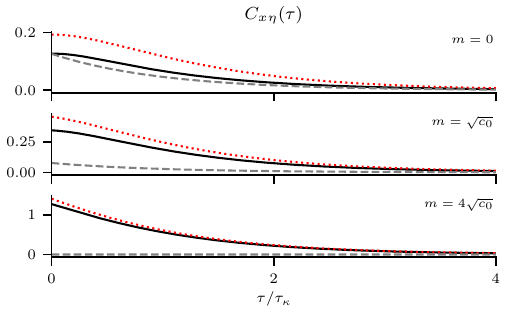}

\caption{\label{fig:Violation-of-Cox}Violation of the CRR for Cox processes
with non-Gaussian intensity due to the non-negativity constraint.
Input-output cross-correlation \prettyref{eq:iocc_linear_nonnegative_cox}
of a linear filter driven by Cox noise with Gaussian intensity that
is clipped at zero (black line). Right-hand side of \prettyref{eq:cox_crr-3}
with the response functions Eqs. \eqref{eq:mean_response_linear_nonnegative_cox}
and \eqref{eq:variance_response_linear_nonnegative_cox} (red dotted
line). The disagreement between red and black lines displays a violation
of the CRR \prettyref{eq:cox_crr-3}. Contribution of the variance-response
to the right-hand side of \prettyref{eq:cox_crr-3}, $2c_{0}K_{xC_{\phi}}(\tau)$
(gray dashed line). Parameters: $\tau_{\kappa}=4$, $\tau_{\phi}=2$,
$c_{0}=0.1$, and $m$ is varied as indicated in the legend.}
\end{figure}

In \prettyref{subsec:cox} we discuss the Cox process, i.e. a conditionally
Poissonian point process $\eta(t)$ with intensity $\lambda(t)$,
and $\lambda(t)$ is itself a random process. Specifically, we choose
$\lambda(t)\equiv\Theta[\phi(t)]\phi(t)$ where $\phi(t)$ is a Gaussian
process with mean $m(t)$ and autocorrelation function $C_{\phi}(t,t^{\prime})$.
In the derivation of the extended form of the CRR for this Cox process,
\prettyref{eq:cox_crr-3}, we assume that $\phi(t)$ is almost always
non-negative, i.e. that the positive mean value is much larger than
the standard deviation. We may thus set $\lambda(t)=\phi(t)$. This
allows us to achieve the closed-form expression of the characteristic
functional $\left\langle \exp\left[\int\lambda[\phi(t)]\left(e^{iu(t)}-1\right)dt\right]\right\rangle _{\phi}$,
see \prettyref{eq:char_func_cox}. Nonlinearities of $\lambda(\phi)$
could be captured systematically by expressing $\lambda$ as a polynomial
around $m$, taking the contributions up to and including $\cO[(\phi-m)^{2}]$
into account by absorbing it into the Gaussian problem, and treating
higher order non-linearities of $\lambda$ using Feynman diagrams.
However, such a procedure is likely to become cumbersome.

Setting $\lambda=\phi$ is legitimate if the probability of negative
$\phi$,
\begin{equation}
\mathds{P}(\phi<0)=\frac{1}{2}\text{erfc}[m(t)/\sqrt{2C_{\phi}(t,t)}],
\end{equation}
can be neglected (here $\text{erfc}$ is the complementary error function).
To estimate the impact of the non-negativity constraint on the validity
of the CRR \prettyref{eq:cox_crr-3}, we first show that \prettyref{eq:cox_crr-3}
is fulfilled for a linear system if indeed $\lambda=\phi$. Second,
we show how \prettyref{eq:cox_crr-3} is violated for a linear system
if $\mathds{P}(\phi<0)\ll1$ is violated.

Assuming $\phi=\lambda$, the response functions of a linear filter
$x[\eta;t]=\int_{0}^{\infty}d\tau\,\kappa(\tau)\eta(t-\tau)$ can
be computed explicitly. Since here, $\left\langle \eta(t)\right\rangle =\left\langle \lambda(t)\right\rangle =m(t)$,
and thus $\left\langle x(t)\right\rangle =\int_{0}^{\infty}d\tau\,\kappa(\tau)m(t-\tau)$,
the linear response functions to mean- and autocorrelation modulations
are
\begin{equation}
K_{xm}(t,t^{\prime})=\frac{\delta}{\delta m(t^{\prime})}\left\langle x(t)\right\rangle =\Theta(t-t^{\prime})\kappa(t-t^{\prime})\label{eq:mean_resp_cox_linear}
\end{equation}
and
\begin{equation}
K_{xC_{\phi}}(t,t^{\prime},t^{\pprime})=\frac{\delta}{\delta C_{\phi}(t^{\prime},t^{\pprime})}\left\langle x(t)\right\rangle =0.
\end{equation}
The fact that $K_{xC_{\phi}}\equiv0$ shows that for $\mathds{P}(\phi<0)\ll1$,
a linear system does not respond to modulations of $C_{\phi}$. 

The input-output cross-correlation can also be computed explicitly
\begin{equation}
\begin{aligned}C_{x\eta}(t,t^{\prime}) & =\int_{0}^{\infty}d\tau\,\kappa(\tau)\left\langle \eta(t-\tau)[\eta(t^{\prime})-m(t^{\prime})]\right\rangle \\
 & =m(t^{\prime})\Theta(t-t^{\prime})\kappa(t-t^{\prime})\\
 & \;+\int dt^{\prime\prime}\,\Theta(t-t^{\pprime})\kappa(t-t^{\pprime})C_{\phi}(t^{\prime},t^{\pprime})\\
 & =m(t^{\prime})K_{xm}(t,t^{\prime})\\
+\int dt^{\pprime} & C_{\phi}(t^{\prime},t^{\pprime})\left[2K_{xC_{\phi}}(t,t^{\prime},t^{\pprime})+K_{xm}(t,t^{\pprime})\right],
\end{aligned}
\label{eq:iocc_cox_apprixmate_negative}
\end{equation}
where in the last step we have recast $C_{x\eta}$ into the form of
the right-hand side of \prettyref{eq:cox_crr-3}, using the above
expressions for the response functions; \prettyref{eq:cox_crr-3}
is thus explicitly confirmed for the linear system. 

If $\mathds{P}(\phi(t)<0)\ll1$ is violated, the above calculations
do not provide a valid approximation anymore. The double-average $\left\langle f[\eta]\right\rangle $,
where first the conditional average over Poisson processes $\eta$
with intensity $\Theta(\phi)\phi$, and then the average over Gaussian
processes $\phi$ have to be carried out, produces expectation values
of a nonlinear function of Gaussian random variables. Yet, the two
response functions can be computed explicitly from the mean output
\begin{equation}
\begin{aligned}\left\langle x(t)\right\rangle  & =\int^{t}dt^{\prime}\,\kappa(t-t^{\prime})\left\langle \eta(t^{\prime})\right\rangle \\
 & =\int^{t}dt^{\prime}\,\kappa(t-t^{\prime})\bar{\eta}[m(t^{\prime}),C_{\phi}(t^{\prime},t^{\prime})]
\end{aligned}
\label{eq:mean_output_cox_linear}
\end{equation}
where
\begin{equation}
\begin{aligned}\bar{\eta}(m,c) & =\left\langle \Theta(\phi)\phi\right\rangle _{\phi\sim\mathcal{N}(m,c)}\\
 & =c\frac{1}{\sqrt{2\pi c}}e^{-m^{2}/(2c)}+\frac{m}{2}\text{erfc}(-m/\sqrt{2c})
\end{aligned}
\end{equation}
is the ensemble average of the input process. The linear response
to mean modulations follows from differentiating \prettyref{eq:mean_output_cox_linear}
\begin{align}
K_{xm} & (t,t^{\prime})=\frac{\delta}{\delta m(t^{\prime})}\int^{t}ds\,\kappa(t-s)\bar{\eta}[m(s),C_{\phi}(s,s)]\nonumber \\
 & =\frac{d}{dh}\int^{t}ds\,\kappa(t-s)\times\label{eq:mean_response_linear_nonnegative_cox}\\
 & \;\,\,\times\bar{\eta}[m(s)+h\delta(s-t^{\prime}),C_{\phi}(s,s)]\Bigg|_{h=0}\nonumber \\
 & =\Theta(t-t^{\prime})\kappa(t-t^{\prime})\frac{1}{2}\text{erfc}[-m(t^{\prime})/\sqrt{2C_{\phi}(t^{\prime},t^{\prime})}].\nonumber 
\end{align}
For $C_{\phi}\rightarrow0$ this reproduces \prettyref{eq:mean_resp_cox_linear},
since $\lim_{x\rightarrow-\infty}\text{erfc}(x)=2$.

Concerning modulations of $C_{\phi}(t,t^{\prime})$, we first observe
that the mean output \prettyref{eq:mean_output_cox_linear} only depends
on the variance $V_{\phi}(t)\equiv C_{\phi}(t,t)$. Thus, the linear
response to modulations of $C_{\phi}$ is given by the linear response
to variance-modulations $K_{xC_{\phi}}(t,t^{\prime},t^{\pprime})=\delta(t^{\prime}-t^{\pprime})K_{xV_{\phi}}(t,t^{\prime})$,
as one can see by Taylor expanding \prettyref{eq:mean_output_cox_linear}
and identifying the correlation response
\begin{align}
\left\langle x(t)\right\rangle _{\epsilon} & =\int^{t}dt^{\prime}\,\kappa(t-t^{\prime})\bar{\eta}[m(t^{\prime}),V_{\phi}(t^{\prime})+\epsilon D(t^{\prime},t^{\prime})]\nonumber \\
 & =\left\langle x(t)\right\rangle _{0}+\epsilon\int^{t}dt^{\prime}\,\kappa(t-t^{\prime})\times\nonumber \\
 & \hfill\qquad\times\frac{\partial}{\partial V_{\phi}(t^{\prime})}\bar{\eta}[m(t^{\prime}),V_{\phi}(t^{\prime})]D(t^{\prime},t^{\prime})\nonumber \\
 & =\left\langle x(t)\right\rangle _{0}+\epsilon\int^{t}dt^{\prime}\int^{t}dt^{\pprime}\,\kappa(t-t^{\prime})\times\\
\times & \delta(t^{\prime}-t^{\pprime})\frac{\partial}{\partial V_{\phi}(t^{\prime})}\bar{\eta}[m(t^{\prime}),V_{\phi}(t^{\prime})]D(t^{\prime},t^{\prime\prime}).\nonumber 
\end{align}
As one may read off from this, the variance-response follows again
by differentiating \prettyref{eq:mean_output_cox_linear}
\begin{align}
K_{xV_{\phi}} & (t,t^{\prime})=\frac{\delta}{\delta V_{\phi}(t^{\prime})}\int^{t}ds\,\kappa(t-s)\bar{\eta}[m(s),V_{\phi}(s)]\nonumber \\
 & =\Theta(t-t^{\prime})\kappa(t-t^{\prime})\frac{1}{2}\frac{1}{\sqrt{2\pi V_{\phi}}}e^{-m^{2}/(2V_{\phi})}.\label{eq:variance_response_linear_nonnegative_cox}
\end{align}
Thus, as opposed to the case where $\mathds{P}(\phi<0)\ll1$, the
nonlinearity induces a non-vanishing response to variance modulations.
This response function vanishes for $V_{\phi}\rightarrow0$, the limit
in which $\mathds{P}(\phi<0)\rightarrow0$, provided that $m>0$.

To test the violation of \prettyref{eq:cox_crr-3}, we additionally
need to compute the input-output cross-correlation
\begin{equation}
\begin{aligned}C_{x\eta}(t,t^{\prime}) & =\int^{t}ds\kappa(t-s)\left\langle \eta(s)\eta(t^{\prime})\right\rangle \\
 & \;\,\,-\int^{t}ds\kappa(t-s)\left\langle \eta(s)\right\rangle \left\langle \eta(t^{\prime})\right\rangle \\
 & =\int^{t}ds\kappa(t-s)\times\\
 & \;\,\,\times\left[\left\langle \lambda[\phi(s)]\delta(s-t^{\prime})\right\rangle +C_{\lambda}(s,t^{\prime})\right]\\
 & =\Theta(t-t^{\prime})\kappa(t-t^{\prime})\bar{\eta}[m(t^{\prime}),V_{\phi}(t^{\prime})]\\
 & \;\,\,+\int^{t}ds\kappa(t-s)C_{\lambda}(s,t^{\prime}).
\end{aligned}
\label{eq:iocc_linear_nonnegative_cox}
\end{equation}
We are not aware of a closed form expression for $C_{\lambda}$ in
terms of $C_{\phi}$, although a possible method is applied in \citep{KruLin16}.
Here, we content ourselves with computing $C_{\lambda}$ numerically
for the special case $C_{\phi}(t,t^{\prime})=c_{0}e^{-|t-t^{\prime}|/\tau_{\phi}}$
by evaluating the double integral
\begin{equation}
\begin{aligned}C_{\lambda}(\tau) & =\int d\phi_{1}d\phi_{2}\mathcal{N}\left[\left.\begin{pmatrix}\phi_{1}\\
\phi_{2}
\end{pmatrix}\right|\begin{pmatrix}m\\
m
\end{pmatrix},\begin{pmatrix}c_{0} & c_{\tau}\\
c_{\tau} & c_{0}
\end{pmatrix}\right]\times\\
 & \,\,\times[\lambda(\phi_{1})-\bar{\eta}(m,c_{0})][\lambda(\phi_{2})-\bar{\eta}(m,c_{0})],
\end{aligned}
\end{equation}
where $c_{\tau}=C_{\phi}(\tau)$, using Gauss-Hermite quadrature.

In \prettyref{fig:Violation-of-Cox} we show the input-output cross-correlation
\prettyref{eq:iocc_linear_nonnegative_cox} as well as the right-hand
side of \prettyref{eq:cox_crr-3} using the response functions Eqs.
\eqref{eq:mean_response_linear_nonnegative_cox} and \eqref{eq:variance_response_linear_nonnegative_cox}
for $\kappa(\tau)=e^{-\tau/\tau_{\kappa}}$ and different values of
the coefficient of variation $\sqrt{c_{0}}/m$. As expected, the CRR
\prettyref{eq:cox_crr-3} is violated if $\sqrt{c_{0}}/m$ is not
small enough, corresponding to $\mathds{P}(\phi<0)\ll1$ being violated.

\end{document}